\documentclass[a4paper,12pt,reqno,superscriptaddress,nofootinbib]{revtex4}
\usepackage[centertags]{amsmath}
\usepackage{amsfonts}
\usepackage{amssymb}
\usepackage{amsthm}
\usepackage{newlfont}
\usepackage{stmaryrd}
\usepackage{mathrsfs}
\usepackage{euscript}
\usepackage{graphicx}
\usepackage{enumerate}
\usepackage{tikz}
\usepackage{pgf}
\usetikzlibrary{arrows,automata,positioning,fit,calc}
\usepackage{wrapfig}
\usepackage{subfigure}
\usepackage{amscd}
\usepackage{hyperref}

\theoremstyle{plain}

\theoremstyle{definition}

\theoremstyle{remark}

\newcommand{\Oe}{{+\mathcal{O}(\varepsilon^3)}}

 \let\be=\beta  
\let\ve=\varepsilon

\newcommand{\caD}{{\mathcal D}}

\newcommand{\caT}{{\mathcal T}}

\newcommand{\bbZ}{{\mathbb Z}}

\newcommand{\opunit}{\text{1}\kern-0.22em\text{l}}

\DeclareMathAlphabet{\mathpzc}{OT1}{pzc}{m}{it}

\newcommand{\id}{\textrm{d}}

\begin{document}

\title{{\bf Non-dissipative effects in nonequilibrium systems }}

\author{Christian Maes\\ {\it Instituut voor Theoretische Fysica, KU Leuven}}

\begin{abstract}
Studying the role of activity parameters and the nature of time-symmetric path-variables constitutes an important part of nonequilibrium physics, so we argue.  The relevant variables are residence times and the undirected traffic between different states.  Parameters are the reactivities, escape rates and accessibilities and how those possibly depend on the imposed driving.  All those count in the {\it frenetic} contribution to statistical forces, response and fluctuations, operational even in the stationary distribution when far enough from equilibrium. As these time-symmetric aspects can vary independently from the entropy production we call the resulting effects non-dissipative, ranking among features of nonequilibrium that have traditionally not been much included in statistical mechanics until recently.  Those effects can be linked to localization such as in negative differential conductivity, in jamming or glassy behavior or in the slowing down of thermalization. Activities may decide the direction of physical currents away from equilibrium, and  the nature of the stationary distribution, including its population inversion, is not as in equilibrium decided by energy-entropy content. The ubiquity of non-dissipative effects and of that frenetic contribution in theoretical considerations invites a more operational understanding and statistical forces outside equilibrium appear to provide such a frenometry. 
\end{abstract}

\maketitle
{ 
\baselineskip=20pt
\section{Introductory comments}\label{int}
Upon opening a book or a review on nonequilibrium physics, if not exposed to specific models, we are often guided immediately to consider notions and quantities that conceptually remain very close to their counterparts in equilibrium and that are concentrating on dissipative aspects.
We mean ideas from local equilibrium, from balance equations and from meditating about the nature of entropy production.  Even in the last decades, while a fluctuation theory for nonequilibrium systems has been moving to the foreground, in the middle stood the fluctuations of the path-dependent entropy fluxes and currents.  A good example of a collection of recent work is stochastic thermodynamics, which however has concentrated mostly on retelling in a path-dependent way the usual thermodynamic relations, concentrating on refinements of the second law and other dissipative features.  Similarly, so called macroscopic fluctuation theory has been restricted to diffusive limits where the driving boundary conditions are treated thermodynamically.  Nevertheless, more and more we see the importance of dynamical activity and time-symmetric features in nonequilibrium situations.  There is even a domain of research now about active particles and active media where the usual driving conditions are replaced by little internal engines or by  contacts with nonequilibrium degrees of freedom and where non-thermodynamic features are emphasized. An important property of these active particles is their persistence length which is of course itself a time-symmetric quantity.   In the present text we call all those the non-dissipative aspects and we will explain in the next section what we exactly mean by that.  Let us however first remind ourselves that the big role of entropic principles in equilibrium statistical mechanics is quite miraculous, and hence should not be exaggerated or tried to be repeated as such also for nonequilibria.\\

For a closed and isolated macroscopic system of many particles undergoing Hamiltonian dynamics one easily identifies a number of conserved quantities such as the total energy $E$, the number $N$ of particles and the volume $V$.  If we know the interaction between the particles and with the walls we can then estimate the phase space volume $W(x;E,V,N)$ corresponding to values $x$ for well-chosen macroscopic quantities $X$ at fixed $(E,V,N)$.  Those $X$ may for example correspond to spatial profiles of particle and momentum density or of kinetic energy etc., in which case the values $x$ are really functions on physical or on one-particle phase space, but in other cases the value(s) of $X$ can also be just numbers like giving the total magnetization fo the system.  At any rate, together they determine what is called the macroscopic condition.  Equilibrium is that condition (with values $x_\text{eq}$) where $W(x;E,V,N)$ is maximal, and the equilibrium entropy is $S_\text{eq} = S(E,V,N) = k_B \,\log W(x_\text{eq};E,V,N)$.  In other words, we find the equilibrium condition by maximizing the entropy functional $S(x;E,V,N) =k_B\log W(x,E,V,N)$ over all possible values $x$.\\
Going to open systems, be it by exchanging energy or particles with the environment or with variable volume, we use other thermodynamic potentials (free energies) but they really just replace for the open (sub)system what the entropy and energy are doing for the total system: via the usual tricks (Legendre transforms) we can move between (equivalent) ensembles.  In particular the Gibbs variational principle determines the equilibrium distribution, and hence gets specified by the interaction and just a few thermodynamic quantities.\\
 Something very remarkable happens on top of all that.  Entropy and these thermodynamic potentials also have an important operational meaning in terms of heat and work. In fact historically, entropy entered as a thermodynamic state function via the Clausius heat theorem, a function of the equilibrium condition whose differential gives the reversible heat over temperature in the instantaneous thermal reservoir.  The statistical interpretation was given only later by Boltzmann, Planck and Einstein, where entropy (thus, specific heat) governs the macroscopic static fluctuations making the relation between probabilities and entropy at fixed energy (which explains the introduction of $k_B$).  The same applies for the relation between e.g. Helmholtz free energy and isothermal work in reversible processes. Moreover, that Boltzmann entropy gives an H-functional, a typically monotone increasing function for the return towards equilibrium.  That relaxation of macroscopic quantities follows gradient flow in a thermodynamic landscape.  Similarly, linear response around equilibrium is related again to that same entropy in the fluctuation--dissipation theorem, where the (Green-)Kubo formula universally correlates the observable under investigation with the excess in entropy flux as caused by the perturbation.  And of course, statistical forces are gradients of thermodynamic potentials with the entropic force being the prime example of the power of numbers. To sum it up, for equilibrium purposes it appears sufficient to use energy-entropy arguments, and in the close-to-equilibrium regime arguments based on the fluctuation--dissipation theorem and on entropy production principles suffice to understand response and relaxation.  All of that is basically unchanged when the states of the system are discrete as for chemical reactions, and in fact much of the formalism below will be applied to that case.\\

Nonequilibrium statistical mechanics wants to create a framework for the understanding of open driven systems. The driving can be caused by the presence of mutually contradicting reservoirs, e.g. holding different temperatures at the ends of a system's boundaries or imposing different chemical potentials at various places.  It can also be implied by mechanical means, like by the presence of non-conservative forces, or by contacts with time-dependent or nonequilibrium environments, or by very long lived special initial conditions. There are indeed a great many nonequilibria, and it is probably naive to think there is a simple unique framework comparable with that of Gibbs in equilibrium for their description and analysis.
It would be rather conservative to believe that extensions involving only notions such as local equilibrium and entropy production, even when space-time variable, would suffice to describe the most interesting nonequilibrium physics.  It is not because stationary non-zero dissipation is almost equivalent with steady nonequilibrium, or that dissipation is ubiquitous in complex phenomena that all nonequilibrium properties would be {\it determined} by the time-antisymmetric fluctuation sector or by energy--entropy considerations, or that typical nonequilibrium features would be uniquely {\it caused} by dissipative aspects.   That is not surprising,  but still it may be useful to get simple reminders of the role of time-symmetric and kinetic aspects in the construction of nonequilibrium statistical mechanics.  The plan of this note is then to list a number of non-dissipative aspects, summarized in what we call the frenetic contribution and related, to discuss the measurability of that.  The point is that non-dissipative features become manifest and visible even in stationary conditions, when sufficiently away from equilibrium --- the (nonequilibrium) dissipation merely makes the constructive role of non-dissipative aspects effective.\\

In these notes we have avoided complications and the examples are kept very simple, just enabling each time to illustrate a specific point.  For example no general physics related to phase transitions or pattern formation is discussed.  Also the level of exposition is introductory.  Yet, the material or the concepts are quite new compared to the traditional line of extending standard thermodynamics to the irreversible domain \cite{Groot}.

\tableofcontents

\section{(Non-)dissipative effects?}
Before we start discussing possible effects and phenomena we need to be more precise about the meaning of dissipative {\it versus} non-dissipative aspects.  As alluded to already in the abstract that plays in two ways: there will be (1) activity parameters, and (2) important time-symmetric path-variables. In general the activity parameters allow more or bigger changes and transitions in the system; we can think how e.g. temperature or diffusion constants allow the system to rapidly explore more state space.  Or how by shaking we can reactivate a cold battery. As examples of time-symmetric variables we can try to observe the sojourn time in a given condition or the undirected traffic between different regions in state space.\\  The easiest way to be more specific about all those is to refer to the modeling via Markov processes, a common tool in nonequilibrium statistical mechanics. For the moment we miss crucial and interesting physics by ignoring spatial extensions or confinements but some important points can (and should) already be illustrated for continuous time jump processes on a finite state space $K$ without insisting on spatial structure or architecture. 
The elements of $K$ are called states $x,y,\ldots \in K$ and can represent the coarse grained position of particle(s) or a chemical-mechanical configuration of a molecule, or an energy level as derived via Fermi Golden's Rule in quantum mechanics etc.  There are transition rates $k(x,y)\geq 0$ for the jump $x\rightarrow y$, and they are supposed to make physical sense of course.  In particular here we have in mind that all such transitions are associated with an entropy flux $s(x,y) = -s(y,x)$ in the environment.  The environment is taken to be time-independent and consisting possibly of multiple equilibrium reservoirs which are characterized primarily by their (constant) temperature or chemical potential.  Their presence in the model is indirect, and the (effective) Markov dynamics should in principle be obtained via some weak coupling limit or other procedures that integrate out the environment.  The point is that the entropy fluxes in these reservoirs are entirely given in terms of the changes in the states of the system. (We no longer call it the open (sub)system from now on.) The $s(x,y)$ is the change of the entropy in one of the equilibrium reservoirs in the environment associated to the change $x\rightarrow y$ in the system.\\ 

In a deep sense that entropy flux $s(x,y)$ measures the time-asymmetry.  The point in general is that we understand our modeling such that the ratio  of transition rates for jumps between states $x$ to $y$ satisfies
\begin{equation}\label{ldb}
\frac{k(x,y)}{k(y,x)} = e^{s(x,y)}
\end{equation}
where $s(x,y) =-s(y,x)$ is the entropy flux per $k_B$ (Boltzmann's constant) over the transition $x\rightarrow y$. That hypothesis \eqref{ldb}, which can be derived in the usual Markov approximation when the reservoirs are well separated, is called the condition of local detailed balance and follows from the dynamical reversibility of standard Hamiltonian mechanics; see \cite{kls,time,har,der,hal}.  It is obviously an important indicator of how to model the time-antisymmetric part of the transition rates.   Loosely speaking here, dissipative is everything which is expressed in terms of those entropy fluxes or other quantities that are anti-symmetric under time-reflection/reversal.  A driving field or force can generate currents with associated entropy fluxes into the various reservoirs in contact with the system.  If we specify a trajectory $\omega = (x_s, s\in [0,t])$ of consecutive states in a time-interval $[0,t]$, then the time-antisymmetric sector contains all functions $J(\omega)$  which are anti-symmetric under time-reversal, $J(\theta\omega) = - J(\omega)$ for $(\theta\omega)_s = x_{t-s}$. We  could for a moment entertain the idea that the nonequilibrium condition of the system would be entirely determined by giving the interactions between the particles and the values of all observables in the time-antisymmetric sector, or even only by the values of some currents or mean entropy fluxes, together with the intensive parameters of the equilibrium reservoirs making up the environment. Or we could hope that the stationary nonequilibrium distribution is determined by a variational principle involving only the expected entropy production as function of probability laws on $K$. All that however would be a dissipative dream, at best holding true for some purposes and approximations close-to-equilibrium.  Non-dissipative effects bring time-symmetric observables to the fore-ground, like the residence times in states or the unoriented traffic between states.  When such observables as the time-symmetric dynamical activity explicitly contribute to the nonequilibrium physics, we will speak of a frenetic contribution.\\

The rates (and hence the modeling) is of course not determined completely by \eqref{ldb}.  We also have the products $\gamma(x,y) = k(x,y)k(y,x) = \gamma(y,x)$ which each are symmetric between forward and backward jumps.  It is like the ``width'' of the transition. Note also that it enters independently from the entropy flux because over all edges where $\gamma(x,y) =\psi^2(x,y)\neq 0$, we can write
\begin{equation}\label{ps}
k(x,y) = \sqrt{k(x,y)k(y,x)}\,\sqrt{\frac{k(x,y)}{k(y,x)}} = \psi(x,y)\,e^{s(x,y)/2}
\end{equation}
We call the $\psi(x,y)=\psi(y,x)\geq 0$ activity parameters; they may depend on the temperature of the reservoir(s) but what is also very important is that they may (as do the $s(x,y)$)  depend on the driving fields, like external forces or differences in reservoir temperatures and chemical potentials.   The $\psi(x,y)$ will be determined again from some weak coupling procedure but can also be obtained from Arrhenius and Kramers type formul{\ae} for reaction rates.  How they will depend on driving (nonequilibrium) parameters is an important challenge.  We count as non-dissipative effect how the $\psi(x,y)$ specify or even determine the nonequilibrium condition, in particular through their variation with the external field.  Again we will speak here about a frenetic contribution.\\

Let us finally compare again with the equilibrium situation.  Here we need the dynamics to be undriven in the sense that the stationary distribution when extended in the time-domain is invariant under time-reversal.  In other words, when under equilibrium we must have that all expectations $\langle J(\omega) \rangle_\text{eq} = 0$ of time-antisymmetric observables $J(\omega)$ vanish.  That is of course much more than requiring stationarity, which only says that $\langle \,f(x_t) - f(x_0)\, \rangle = 0$ for all times $t$.   Time-reversal invariance in the stationary condition (reversibility or equilibrium, for short)
is equivalent with having \eqref{ldb} for $s(x,y) = {\cal F}(x) -{\cal F}(y)$ for some free energy function $\cal F$ on $K$.  We do not prove that statement here, but the reader then recognizes the typical expressions for transition rates under (global) detailed balance, as
\[
k(x,y) = \psi(x,y)\,\exp[{\cal F}(x) - {\cal F}(y)]/2,\qquad \psi(x,y)=\psi(y,x)
\]
with stationary distribution $\rho_\text{eq}(x) \propto \exp-{\cal F}(x)$ as prescribed by Gibbs.  Note that $\rho_\text{eq}$ does not depend on the activity parameters $\psi(x,y)>0$; there is no such frenetic contribution in equilibrium.\\

  We start in the next section with non-dissipative effects on the stationary distribution and then we go on with other instances for the current, in response etc.  Beyond and above these examples it  should be clear however that as such dynamical activity is present as an important time-symmetric background for systems even before perturbations or other changes are applied.
In some way we find in it the analogue of a {\it vis viva} through which typical nonequilibrium phenomena can get realized.  Taking now living matter indeed, it has for example become clear in the last decade
``that stimulus- or task-evoked activity accounts for only a fraction of the metabolic budget of the brain, and that intrinsic activity, i.e. not stimulus- or task-driven activity, plays a critical role
in brain function'' \cite{cor}. There is also the older ``vacuum activity'' coined by Konrad Lorenz in the 1930's, for innate patterns of animal behaviour that are there even in the absence of external stimuli.  That has nothing to do with ``vacuum polarization'' or the self-energy of the photon but that in itself is a dynamical activity of the vacuum which is crucial for electrodynamics in the quantum regime and will change under nonequilibrium; see e.g. \cite{luit}.

\section{On the stationary distribution}\label{onst}

For the continuous time Markov processes we are concentrating on now, there is a unique stationary distribution $\rho$ which is reached exponentially fast in time and uniformly so over all initial conditions.
We assume in other words irreducibility of the Markov process over the finite state space $K$ so that all nodes are connected via some path on the graph having the states as vertices and with edges over pairs $\{x,y\}$ where $\psi(x,y)\neq 0$.\\
The backward generator is
$(L f)(x) := \sum_y k(x,y)\,[f(y) - f(x)]$, and the unique stationary probability distribution $\rho>0$ on $K$ is characterized by  $\sum_x \rho(x) (Lf)(x) =0$ for all functions $f$, which means $\sum_y [\rho(x)k(x,y) - \rho(y)k(y,x)] = 0$ for all $x$ (stationary Master equation).  The stationary distribution $\rho$ will depend on the parameters in \eqref{ps}, and we want to highlight the frenetic contribution.
Here is the simplest example.\\

Consider two cells which can each be either vacant or occupied (by one particle), in contact with equilibrium particle reservoirs. We have in mind that particles can enter the system from both reservoirs, say form left and from right, and similarly that particles can leave the system by exiting to the left or to the right reservoir. There can be at most one particle per cell, as in the Coulomb blockade regime of quantum (double) dots.  The model dynamics is a Markov process with states in $K= \{00, 01,10,11\}$ where each element (state)  refers to a vacancy $0$ or an occupation $1$ for each of the two cells. The transition rates are 
\begin{eqnarray*}
	k(01,10) =1,& \quad & k(10,01) = 1 \nonumber\\
	k(0*,1*) = \alpha, \quad k(1*,0*) = \gamma,&& \quad k(*0,*1) = \delta, \quad k(*1,*0) = \kappa \nonumber 
\end{eqnarray*}
where $*=0,1$ is to be substituted; e.g. $k(01,11) = k(00,10) = \alpha$.  The first line is a simple exchange of occupations which happens dissipationless, $s(01,10) = s(10,01) = 0$.  The second line is about the transitions where a particle enters or exits the first or the second cell; then the occupation either in the first or in the second cell switches.\\
The parameters $\alpha$ and $\delta$ are entrance rates, $\gamma$ and $\kappa$ are exit rates, to the left and to the right particle reservoirs respectively. The thermodynamics enters in the ratios $\alpha/\gamma =\exp \mu_\ell$  and $\delta/\kappa =\exp \mu_r$, determined by the chemical potentials $\mu_\ell$ and $\mu_r$ characterizing the left and right reservoir respectively (at inverse temperature $\beta=1$).  That is an example of requiring {\it local detailed balance} \eqref{ldb} for the transition rates.  The stationary distribution, which gives the histogram of occupations for many repeated observations, can be computed from the stationary Master equation,
\begin{eqnarray*}
	\text{Prob}[00] &=& ( \kappa^2 + \gamma^2 (1 + \kappa) + \gamma \kappa (2 + \alpha + \delta + \kappa) )/z\\
	\text{Prob}[01] &=&(\alpha (\gamma + \delta \gamma + \kappa) + \delta (\kappa + \gamma (1 + \delta + \gamma + \kappa)) )/z\\
	\text{Prob}[10]&=& ((\alpha + \delta) \gamma + (\delta + \alpha (1 + \alpha + \delta + \gamma)) \kappa + \alpha \kappa^2)/z\\ 
	\text{Prob}[11]&=& (\delta^2 + \alpha^2 (1 + \delta) + \alpha \delta (2 + \delta + \gamma + \kappa))/z
\end{eqnarray*}
for normalization $z:=(\alpha + \delta + \gamma + \kappa) (\delta + \gamma + \kappa + \gamma (\delta + \kappa) + \alpha (1 + \delta + \kappa))$.
Clearly, from symmetries and from normalization, such distribution is completely decided from knowing
\begin{equation}\label{B}
\text{Prob}[01]/\text{Prob}[00] =: B
\end{equation}
as a function of $(\alpha,\gamma,\delta,\kappa)$.

\begin{figure}[h]
	\centering
	\includegraphics[width=15 cm]{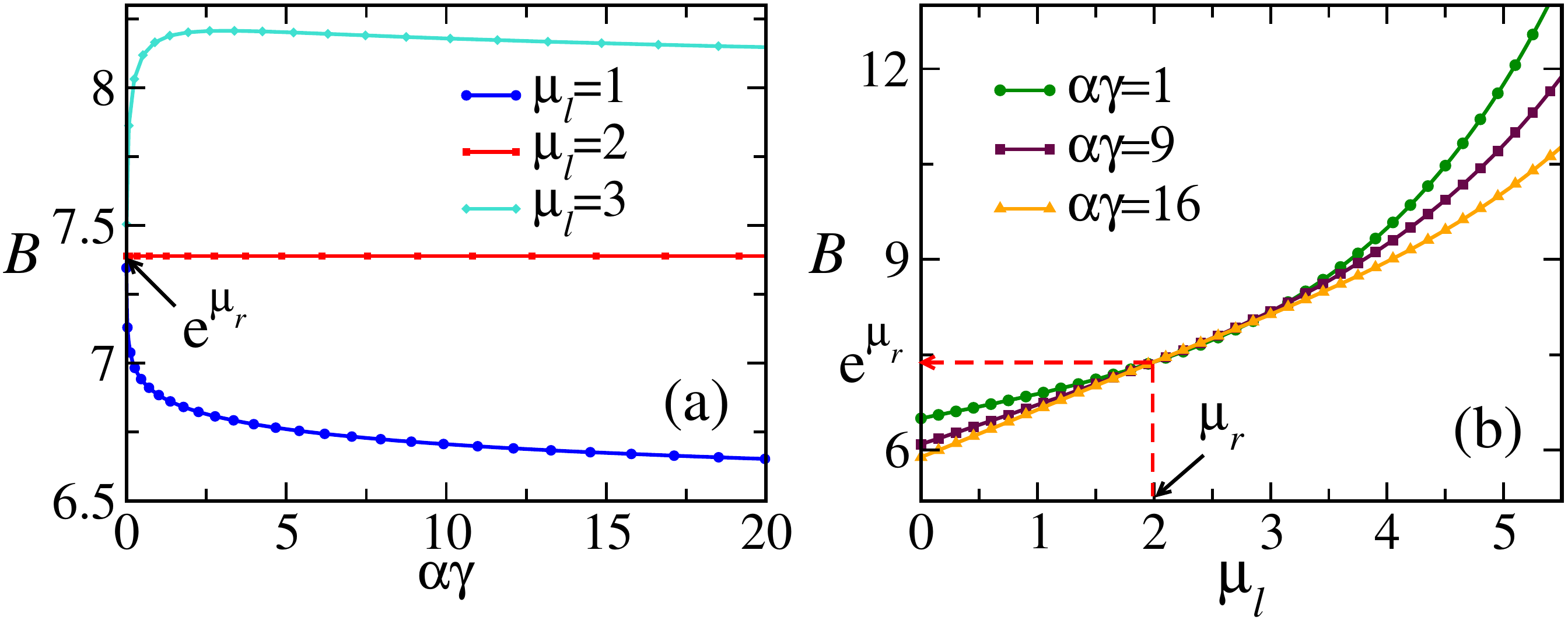}
	\caption{The driven two-cell system for which the ratio $B$ is defined in \eqref{B}. When fixing the densities/chemical potentials of the baths, the kinetics in terms of the entry/exit rates, here represented by the product $\alpha \gamma$, changes $B$ when $\mu_\ell\neq \mu_r.$ }
	\label{4state}
\end{figure}

Under equilibrium (no chemical driving) at chemical potential $\mu_\ell =\mu =\mu_r$ (same for left and right reservoirs), when $\alpha\kappa=\gamma\delta$, the expression for the stationary distribution gives $B= \exp \beta \mu$ as it should for the grand-canonical distribution, completely determined by the thermodynamic specification of the environment.  Not so when $\mu_\ell\neq \mu_r$, i.e., under nonequilibrium driving conditions: even when fixing $\mu_\ell$ and $\mu_r$, but changing say the product $\alpha\gamma$ we get different values for the ratio $B$, see Fig.~\ref{4state}.  In other words, the stationary distribution is not thermodynamically specified.  The activation parameters $\alpha\gamma$ and $\delta\kappa$ matter truly. There is really not much surprising here, and the reader can easily construct her own examples. Yet, in that way it all remains rather anecdotal.  One of the major challenges of nonequilibrium statistical mechanics is to bring order in that situtation or in those probability laws --- what is the best parameterization of the stationary distribution for a physically relevant characterization?  How to collect the most important non-dissipative aspects that determine the stationary distribution?\\

The above example is easy but stands for the general scenario that static fluctuations in the stationary distribution far from equilibrium will depend on activity parameters, and in contrast with equilibrium, are not directly related to thermodynamics like via energy-entropy like quantities. Still, when the model above is extended to the boundary driven exclusion process, and one looks in the diffusive scaling limit (of space-time) one finds nonequilibrium free energies which retain important nonequilibrium characteristics but the dependence on the kinetics is gone.  Clearly the precise formulae will be difficult to get analytically\footnote{But there is some algorithm, where the static fluctuation functional becomes the solution of a Hamilton-Jacobi equation, see \cite{Japan,macro}.  In equilibrium we have the macroscopic static fluctuation theory of Boltzmann-Planck-Einstein. It is still very instructive to read the first pages of \cite{einstein} to get an early review.  Later reviews for the macroscopic fluctuation theory in equilibrium are for example \cite{lanford, martinlof,ellis}.}, but the general structure for the probability of a density profile is
\begin{equation}\label{nefe}
\text{Prob}_N[\rho(r), r\in [0,1]] \simeq \exp(-N I[\rho(r), r\in [0,1]])
\end{equation}
where $I$ is called the nonequilibrium free energy and $N$ is the rescaling parameter (e.g. the size of the system in microscopic units). Here it is for the open symmetric exclusion process, \cite{derr,matrixprod}, just for completeness,
\[
I[\rho(r), r\in [0,1]] = \int_0^1 \id r\left[\rho(r)\log\frac{\rho(r)}{F(r)} + (1-\rho(r))\log\frac{1-\rho(r)}{1-F(r)} + \log \frac{F'(r)}{\rho_1-\rho_0}\right]
\]
where the function $F$ is the unique solution of
\[
\rho(r)= F(r) + \frac{F(r)(1-F(r))F''(r)}{(F'(r))^2}, \quad F(0)=\rho_0,\,\, F(1)=\rho_1
\]
Note that on that hydrodynamic scale the fluctuations indeed do not appear to depend on the details of the kinetics in terms of the exit and entry parameters at the edges; only the densities $\rho_0,\rho_1$ of the left and right particle reservoirs count in the fluctuation functional $I$, no matter whether the reservoirs contain champagne or water.  Yet that disappearing of the relevant role of activity parameters is restricted to the diffusive regime and to the macroscopic fluctuation theory we alluded to in the beginning of the introduction of these notes.

\subsection{The difference between a lake and a river}

We mentioned above that for (global) detailed balance the equilibrium distribution $\rho_\text{eq}$ does not depend on the activity parameters $\psi(x,y)$.  In fact, in equilibrium if all states in $K$ remain connected via some transition path, adding kinetic constraints (i.e., imposing $\psi(x,y)=0$ for some $(x,y)$) has no effect on the stationary distribution; we retain $\rho_\text{eq}(x) \propto \exp-{\cal F}(x)$. Leaving even aside the required irreducibility, by isolating parts of the state space which are no longer accessible when not starting in it, locally the stationary distribution really does not change drastically.  For example we can look at the dynamics on fewer states, like restricting to smaller volume and ask how the stationary distributions resemble.  The answer is well known, under detailed balance  the stationary distribution for the smaller volume (on a restricted class of states) is the restriction of the equilibrium distribution on the larger volume (original state space), possibly with some changes at the boundary only.  In other words, if a big lake gets perturbed by a wall in the middle, there just appear two (very similar) lakes.\\

In nonequilibrium, setting some activity parameters $\psi(x,y)$ to zero can have drastic effects.  For example, suppose we have a ring $K=\bbZ_N$, $x=1,2,\ldots, N$ with $N+1=1$, with transition rates
$k(x,x+1) = p, k(x,x-1)=q$ (random walker on a ring).  Then the uniform distribution $\rho(x) = 1/N$ is invariant for all values of $p$ and $q$.  Let us now break the ring by putting $k(1,N)=0=k(N,1)$. The dynamics remains well defined and irreducible on the same state space $K$ but now the stationary distribution is
\[
\rho(x) \propto \left(\frac{p}{q}\right)^x
\]
and {\it only} in equilibrium, for $p=q$, is the stationary distribution (unchanged) uniform.  For nonequilibrium, at driving $\log p/q\neq 0$, the uniform distribution has changed into a spatially exponential
profile.  Throwing a tree or building a wall in a river has a much larger effect than for a lake. The river can even turn into a lake.\\   In the next section we give an example how you can use the activity parameters to {\it select} one or more states for the stationary distribution to be concentrated on.

\subsection{From the uniform to a peaked distribution}

Suppose a three-state Markov process with state space $K=\{1,2,3\}$ and transition rates,
\begin{eqnarray}
k(1,2) = a\,e^{\varepsilon/2},\quad k(2,3) = b\,e^{\varepsilon/2},\quad k(3,1) = c\,e^{\varepsilon/2},\nonumber\\
k(1,3) = c\,e^{-\varepsilon/2},\quad k(3,2) = b\,e^{-\varepsilon/2},\quad k(2,1)=a\,e^{-\varepsilon/2}\label{str}
\end{eqnarray}
parameterized by the external field $\varepsilon\geq 0$ and activity parameters $0< a<b\leq c$; see Fig.~\ref{Teff}(b).


If $\varepsilon=0$, there is detailed balance with the equilibrium distribution being uniform on $\{1,2,3\}$, whatever the $a,b,c$.  From the moment $\varepsilon>0$, the asymptotic behavior is that of steady nonequilibrium where the driving $\varepsilon$ does not as such distinguish between the three states.  However, the prefactors $a,b,c$ (activity parameters), while symmetric over the jumps $1\leftrightarrow 2\leftrightarrow 3$,  now determine the stationary condition as illustrated in Fig.~\ref{3st}(b).  Moreover and as is easy to understand, for large $\varepsilon$ the stationary distribution concentrates on that state from which the escape rate is minimal; see Fig.~\ref{3st}(a).
That is an instance of what is sometimes called the Landauer blowtorch theorem \cite{land,heatbounds} on which we say more in the next section.

\begin{figure}[h]
	\centering
	\includegraphics[width= 13 cm]{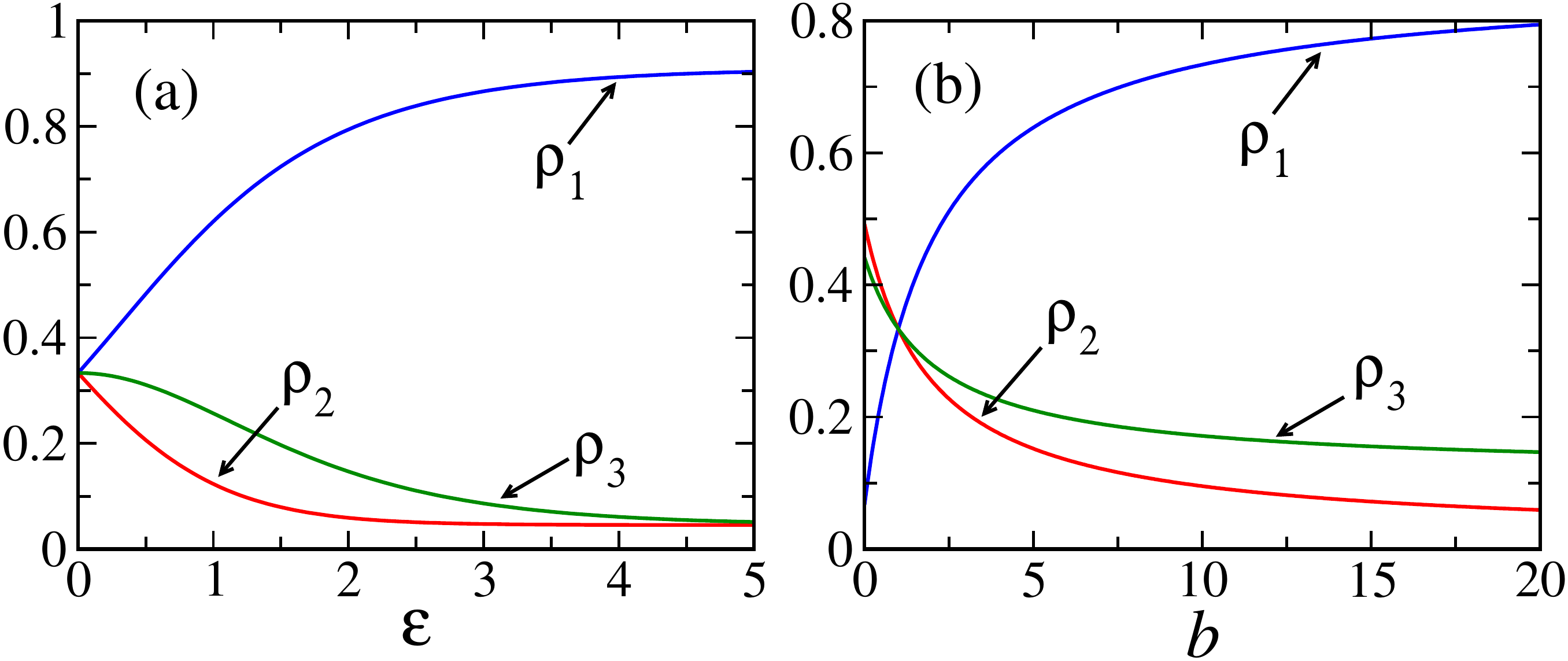}
	\caption{(a) Stationary occupations in the 3-state Markov process as function of $\varepsilon$ for the choice $a=1, b=c=20$.  (b) For fixed $\varepsilon=2$ the stationary occupations as function of $b=c$ for $a=1$.}
	\label{3st}
\end{figure}

That kinetic selection of a particular state is probably at work in a variety of bio-chemical systems for getting a more reliable reproduction or copy of certain proteins, cf. the idea of kinetic proofreading \cite{hopf} and biological error correction.  It illustrates a time-symmetric aspect, non-dissipative, but of course it works only out-of-equilibrium to select a state or configuration.

\subsection{Heat bounds}\label{heab}
Let us go back to \eqref{ldb} but now write the entropy flux as heat over environment temperature,
\begin{equation}\label{tra}
k(x,y) := \psi(x,y)\,\exp \Bigl[ \frac{\beta}{2} q(x,y) \Bigr]
\end{equation}
where we took $\beta= (k_BT)^{-1}\geq 0$ for the inverse temperature of the environment so that the heat is $q(x,y) = -q(y,x)$ following the prescription of local detailed balance \eqref{ldb}, $s(x,y) = \beta q(x,y)$.  The activity parameters are $\psi(x,y) = \psi(y,x) \geq 0$. The edges of $K$ (as a graph) are made by the pairs $\{x,y\}$ over which $\psi(x,y) > 0$.\\
  We want to understand the stationary occupations $\rho(x)$ in terms of the heat $\{q(x,y)\}$ and activity parameters $\{\psi(x,y)\}$.\\

For any oriented path $\caD$ along the edges $b=(x,y)$, let $q(\caD)$ be the total dissipated heat along $\caD$, which is the sum
$q(\caD) := \sum_b q(b)$. For a
spanning tree $\caT$ we put $\caT_{xy}$ for the unique oriented path from $x$ to $y$ along the edges of $\caT$.  Then, we show in \cite{heatbounds} that the stationary occupations satisfy
\begin{equation}\label{occupation-bounds}
\min_{y \stackrel{\!\!\caD}{\rightarrow x}} q(\caD) \leq
\frac{1}{\be} \log\frac{\rho(x)}{\rho(y)} \leq
\max_{y \stackrel{\!\!\caD}{\rightarrow x}} q(\caD)
\end{equation}
with the minimum and maximum taken over all oriented paths (self-avoiding walks) from $y$ to $x$ on the graph.  In the case of \emph{global} detailed balance $q(x,y) = {\cal F}(x)- {\cal F}(y)$ we have that the heat $q(\caD) = {\cal F}(x_i) - {\cal F}(x_f)$ only depends on the initial and final configurations $x_i,x_f$ of the path $\caD$.  We then get the Boltzmann equilibrium statistics for $\rho$.  However in nonequilibrium systems, most of the time it is easy to find configurations $x$ and $y$  for which there exist two oriented paths $\caD_{1,2}$ from $x$ to $y$ such that $q(\caD_1) < 0 < q(\caD_2)$ (heat-incomparable).  In other words along one path ${\cal D}_2$ heat is dissipated while transiting from $x$ to $y$, while along the other path ${\cal D}_1$ heat is absorbed to go from $x$ to $y$.  Then, it simply cannot be decided which of the occupations
$\rho(x)$, $\rho(y)$ is larger on the basis of  heat functions $q(x,y)$; we only have the heat bounds \eqref{occupation-bounds} and nothing more can be concluded from dissipative characteristics.  The activity parameters then become essential.
In such a case indeed where a pair of states $x^*$ and $y^*$ is ``heat-incomparable'', then without changing the heat function $\{q(x,y)\}$, we can  always make either $\rho(x^*) > \rho(y^*)$ or
$\rho(x^*) < \rho(y^*)$ by just changing the $\{\psi(x,y)\}$.   That is typical in nonequilibrium systems; we cannot (partially) order states depending on whether heat is being released over all paths connecting them, or being absorbed over all paths connecting them.  In such a case, when depending on what path we choose heat is absorbed or released in going $x\rightarrow y$, then we can change the relative weight $\rho(x)/\rho(y)$ in the stationary distribution $\rho$ from being greater than one to being smaller than one, just by an appropriate change in the activity parameters (the $a, b, c$ in the previous section).  To all that we can add that the heat bounds \eqref{occupation-bounds} become less sharp of course when the environment temperature lowers (high $\beta$).  We can then expect that the notion of ground state for nonequilibrium systems is not at all (only) energy-connected but also takes into account non-dissipative aspects such as accessibility; see \cite{lowT,win}.

\subsection{Population inversion}

Thermal equilibrium occupation statistics is completely determined by energy-entropy considerations.  In the simplest case of a nondegenerate multilevel system, we have relative occupations determined by temperature and energy difference. However, when adding kinetic effects, like introducing a symmetrizer between two levels we break detailed balance and we can basically select any desired (even high) energy level to have the largest occupation.  In that way, like for lasers, under nonequilibrium conditions but via changes in non-dissipative factors, we can  establish an inversion of the population with respect to the usual Boltzmann statistics.\\

Consider a system with $K$ energy levels where the lowest and highest levels are connected by an equalizer and an additional energy barrier $F$ exists between the $K$ and $K-1^{th}$ level. Denote the number of particles at level $\ell$ by $n_\ell$ and let the rate at which a single particle jumps from level $\ell$ to level $\ell'$ be $k(\ell,\ell')$ with
\begin{eqnarray}
k(\ell,\ell+1 ) &=& n_{\ell}~ e^{-\frac 1{T} (E_{\ell+1} -E_\ell)}, \quad \forall \ell \ne K-1, K \cr
k(\ell,\ell-1) &=& n_{\ell}, \qquad \qquad \quad \forall \ell \ne K,1 \cr
k(K-1,K ) &=& n_{K-1}~ e^{-\frac F{T}} e^{-\frac 1{T} (E_{K} -E_{K-1})},\;\qquad k(1,K) = b~ n_{1}  \cr
k(K,K-1 ) &=& n_{K}~ e^{-\frac F{T}},\;\qquad \qquad k(K,1) = b~ n_{K}
\label{eq:zrp_dyn}
\end{eqnarray}
The equalizing symmetric activity parameter $b>0$ between the highest and lowest energy level along with the factor $e^{-F/T}$  gives rise to the desired population inversion \cite{lowT}.  That can be witnessed  by the ``effective temperature''
\begin{eqnarray}\label{eft}
T_\text{eff} = (E_K -E_1) \left[\log \frac{\rho_1}{\rho_K}\right]^{-1}
\end{eqnarray}
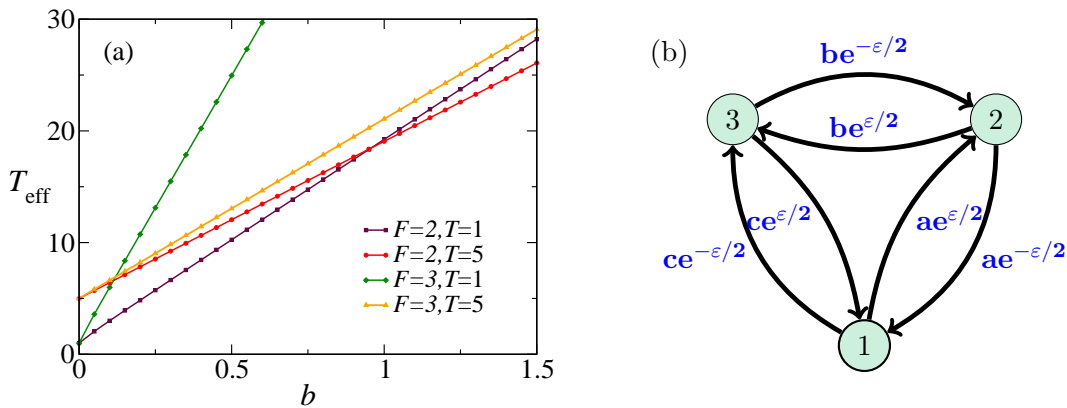
\begin{figure}[h]
	\centering
	\includegraphics[width=7.2 cm]{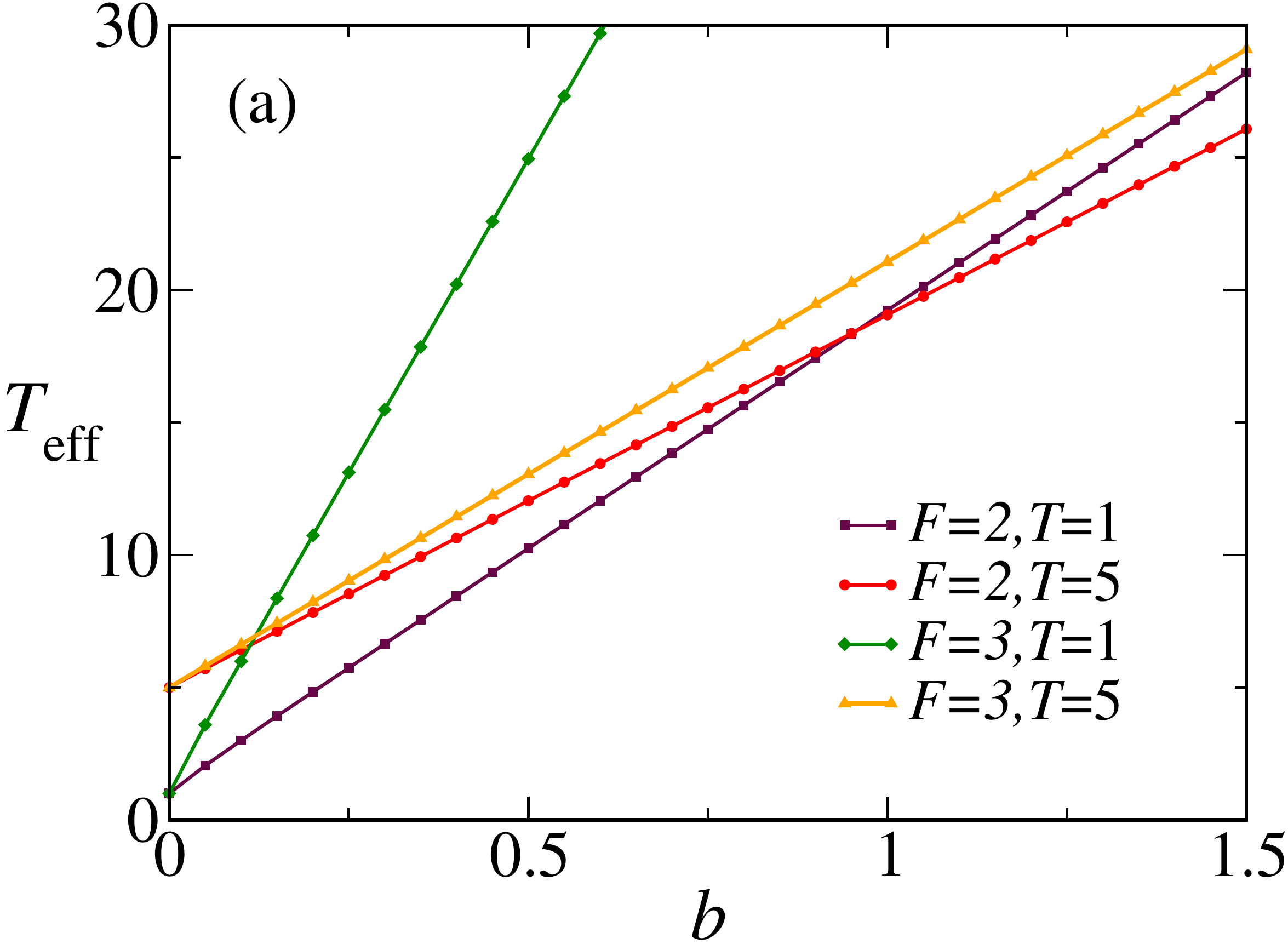}\hspace*{1.0 cm}
	\begin{tikzpicture}
	\node [draw, shape= circle,fill=blue!30!green!20,thick,label=90:$ $](1) at (270:2)  {$1$};
	\node [draw, shape= circle,fill=blue!30!green!20,label=135:$ $] (2) at (30 :2) {$2$};
	\node  [draw, shape= circle,fill=blue!30!green!20,label=180:$ $ ](3) at (150:2) {$3$};
	\node[](4) [above left=0.3 cm and 0.2 cm of 3]{(b)};
	\node[](4) [below left=3.3 cm and 0.2 cm of 3]{};
	\path[every node/.style={font=\sffamily\small}]
	(1) edge  [bend left=20, ->,line width=1.8 pt] node [right] {\color{blue}$\mathbf{a e^{\varepsilon/2}}$} (2)
	(2) edge  [bend left, ->,line width=1.8 pt] node [right ] {\color{blue}$\mathbf{a e^{-\varepsilon/2}}$} (1)
	(1) edge  [bend left, ->,line width=1.8 pt] node [left]{\color{blue}$\mathbf{c e^{-\varepsilon/2}}$} (3)
	(3) edge  [bend left=20, ->,line width=1.8 pt] node [left] {\color{blue}$\mathbf{c e^{\varepsilon/2}}$} (1)
	(2) edge  [bend left=20, ->,line width=1.8 pt] node [above]{\color{blue}$\mathbf{b e^{\varepsilon/2}}$} (3)
	(3) edge  [bend left,->,line width=1.8 pt] node [above]{{\color{blue}$\mathbf{be^{-\varepsilon/2}}$}} (2);
	\end{tikzpicture}
	\caption{(a) Effective temperature $T_\text{eff}$ as function of the symmetrizer $b$ for different environment temperatures $T$ and barrier strengths $F$ for $K=3$ levels.  For $b=0$ there is no dependence on $F$.  For lower $T$ and $b>0$ the dependence on $F$ shows more.  (b) Graph-representation of the 3-state Markov process \eqref{str}.  Changing the activity parameters $a,b,c$ can select a state when $\varepsilon>0$ is big enough.}
	\label{Teff}
\end{figure}

Fig.~\ref{Teff}(a) (a variation of Fig.~7(a) in \cite{kolk}) shows that the effective temperature $T_\text{eff}$ is increasing with the strength of the equalizer $b$.  At low temperature $T$ the effective temperature $T_\text{eff}$ will grow fast with $b$ (and for high $T$ the effective $T_\text{eff}$ is about constant).  That shows up in the figure: for a fixed $F,$ the curve of $T_\text{eff}$ corresponding to a lower thermodynamic temperature $T$ crosses that of a higher temperature from below.  That signifies population inversion as function of the activity parameter $b$ for low temperature $T$. The barrier $F$ (again time-symmetric) facilitates that phenomenon --- crossing occurs for smaller $b$ when $F$ is increased.

\subsection{Variational principles}\label{vari}
Static fluctuations refer to the occurrence of single time events that are not typical for an existing condition.  For example, in this room there is a certain air density which is quite homogeneous and as a result we have a constant index of refraction of light etc.  That is typical for the present equilibrium condition here for air at room temperature and at atmospheric pressure.  Yet, there are fluctuations, meaning little regions where the density locally deviates.  We can expect these regions to be very small; otherwise we would have noticed before.  That is why such fluctuations are called {\it large} deviations; they are very unlikely when they take macroscopic proportions and they are exponentially small in the volume of the fluctuation.  The rate of fluctuations, i.e., what multiplies that volume, is given by the appropriate free energy difference (at least in equilibrium); for the local air density we would have the grand canonical potential. Similarly, the energy $E_V$ in a subvolume $V$ fluctuates.  The total energy is conserved in an isolated system, but there will be exchanges through the boundary; see Fig.~\ref{can}. It could be that the particle number $N$ is also fixed inside $V$, in which case the fluctuations are governed by the Helmholtz free energy $\cal F$, in the following sense.   When the system is in thermal equilibrium and away from phase coexistence regimes\footnote{Otherwise, we must introduce also surface tensions and the scaling could be with the surface of the subsystem and not with its volume.} the local fluctuations of the energy density $E_V/V$
satisfy the asymptotic law
\begin{equation}\label{few}
\text{Prob}\left[\frac{E_V}{V} = e \right] \simeq \exp-\beta V[\cal F(e)-{\cal F}_\text{eq}]
\end{equation}
where the variational functional is ${\cal F}(e) = e - T S(e,V,N)$ with $S$ being the entropy density at that energy density $e$, and ${\cal F}_\text{eq}(T,V,N)$ being the equilibrium free energy density at temperature $T$ (and $\beta^{-1}=k_BT$).

\begin{figure}[h]
	\centering
	\includegraphics[width= 8 cm]{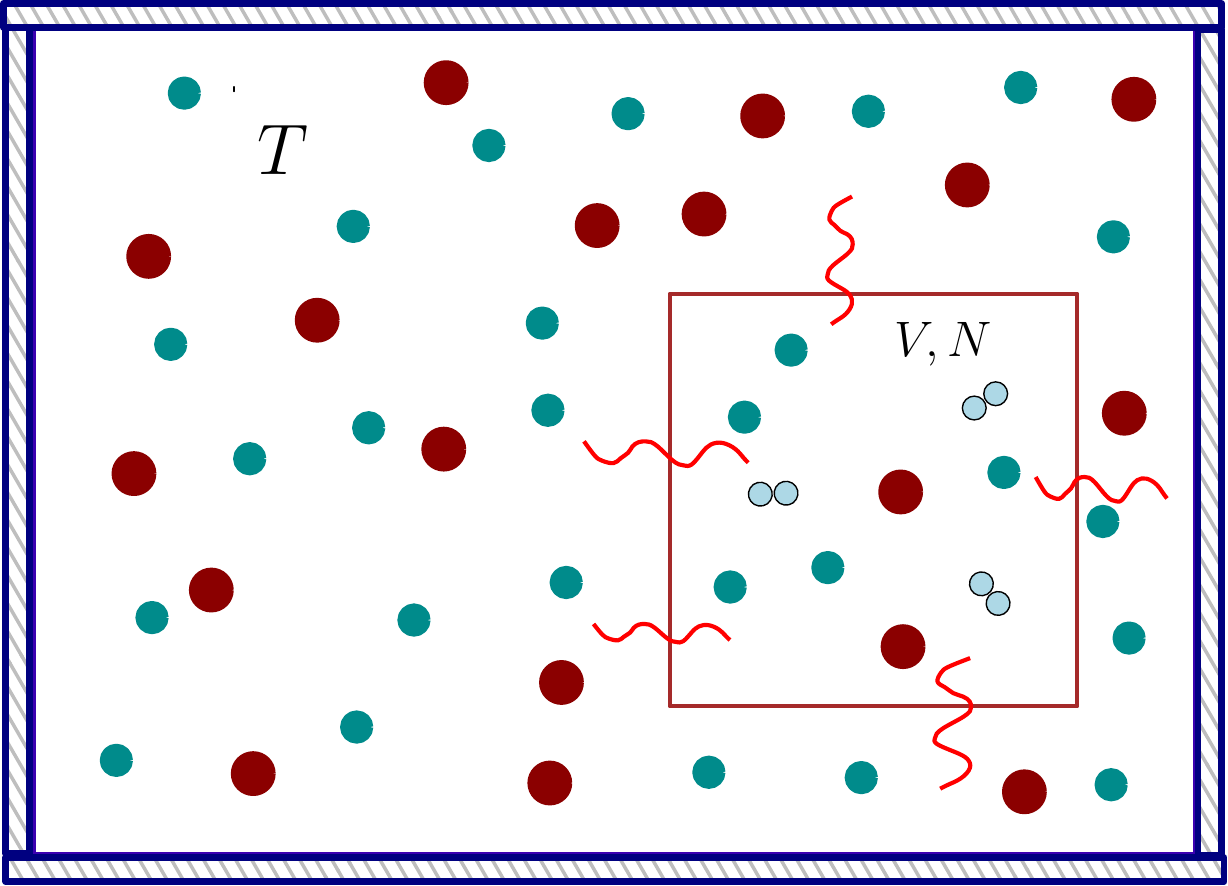}
	\caption{Fluctuations in a subvolume $V$ in weak contact with an equilibrium bath at fixed temperature $T$ are described in the canonical Gibbs ensemble.}\label{can}
\end{figure}

It is only a slight variation to ask for the probability of profiles, i.e., to estimate the plausibility of given spatial variations of a particular quantity.  We then need spatially extended systems, most interestingly with interacting components and that is what we already did around formula \eqref{nefe}.\\

As a clear consequence of  fluctuation formul{\ae} as \eqref{few} we get that the equilibrium condition minimizes the free energy functional, ${\cal F}(e)$ in the above. That constitutes in fact a derivation of the Gibbs variational principle. As such, one could try to repeat that for nonequilibrium systems, but clearly such a formulation with explicit functionals does not exist (yet).  Since a long time however people have been trying to use a dissipative characterization of the stationary distribution.  These are known as entropy production principles; see the minimum entropy production principle discussed in \cite{schEP}.\\
The entropy production rate functional corresponding to the Markov process characterized by \eqref{ps} is
\begin{equation}\label{sig}
\sigma(\mu) = \sum_{x,y}\mu(x) k(x,y)\, \log \frac{k(x,y)\mu(x)}{k(y,x)\mu(y)} \geq 0
\end{equation}
defined on all (possibly even unnormalized) distributions $\mu$.  That functional $\sigma$ is convex, and  homogeneous, $\sigma(\lambda \mu) = \lambda \,\sigma(\mu)$.  In our case of irreducible Markov processes it is even strictly convex.  It thus has a unique minimum, called the Prigogine distribution $\rho_P>0$.  Suppose now that $k(x,y) = k_\varepsilon(x,y)$ depends on a driving parameter $\varepsilon$ so that there is detailed balance for $\varepsilon=0$, $k_0(x,y) = k_\text{eq}(x,y)$, with smooth dependence on $\varepsilon$ close to equilibrium.  Then,  the stationary distribution $\rho^{(\varepsilon)}$ and the Prigogine distribution $\rho_P^{(\varepsilon)}$ coincide up to linear order in $\varepsilon$: $\rho_P^{(\varepsilon)} = \rho^{(\varepsilon)} + O(\varepsilon^2)$.  That is called the minimum entropy production principle (here formulated for finite state space Markov jump processes).  It should be added that most often, the Prigogine distribution as completely characterized by minimizing the entropy production rate \eqref{sig} is of course {\it not} equal to the (true) stationary distribution and they really start to differ from second order onwards.  The reason is just a non-dissipative effect.\\
For a discussion on maximum entropy production principles, we refer to \cite{lincu}.

\subsection{Recent examples}
\subsubsection{Demixing}

The above examples are extremely simple, but the heuristics can easily be moved towards more interesting applications.  Suppose indeed that we have a macroscopic system with two types of particles and we must see whether a condition with phase separation between the two types is most plausible.
In equilibrium that would be called a low-entropy condition which can only be obtained at sufficiently low temperature and with the appropriate interactions.  In nonequilibrium opens the possibility of a totally different physics, that the demixed configuration gets more plausible whenever it is a trap in the sense that the escape rates to leave from it are rather low.  For that to be effective, we need, as above in Section \ref{heab}, that the mixed and the demixed condition are dynamically connected through both  positive dissipative as well as negative dissipative paths.  In the end it will be the configuration with lowest escape possibilities that will dominate.\\
An example of that phenomenon is shown in \cite{frey,ioanny}.  The importance of life-time considerations especially at low temperatures is discussed in \cite{lowT}.

\subsubsection{No thermodynamic pressure}

Suppose we have active particles in a container, like for active Brownians or for self-propelled particles etc.
The pressure on a wall is obtained from calculating the mechanical force on the wall and to average over the stationary ensemble.
Since that stationary ensemble could depend on kinetic details, we cannot expect the pressure to be thermodynamically determined.  It means that details of the interaction between particles and wall can matter and that, unless we have symmetries that cancel the kinetic dependencies, we will not have an equation of state relating that wall pressure to bulk properties such as global density or temperature.  The simple reason is that the stationary distribution is itself not thermodynamically energy-entropy characterized.\\
We find an analysis of that effect in \cite{solon}.

\section{Transport properties}

Transport is usefully characterized in terms of response coefficients such as conductivities.  We discuss them in Section \ref{reps}, while here we deal with the question of what determines the direction of the current and how it could decrease by pushing harder.

\subsection{Current direction decided by time-symmetric factors} 
Consider the metal rod in Fig.~\ref{hecs} which is connected at its ends with a hot and a cold thermal bath.  The environment exchanges energy with the system but at different temperatures for the two ends of the rod. 

\begin{figure}[h]
\centering
\includegraphics[width=8 cm]{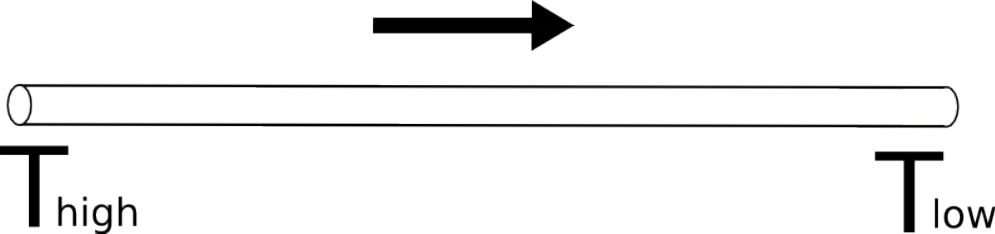}
\caption{Example of a simple stationary current for which the direction is decided by the positivity of the entropy production.}\label{hecs}
\end{figure}

 In that case we can, on the appropriate scale of time, think of the system as being in steady nonequilibrium.  It is stationary alright, not changing its macroscopic appearance, but a current is maintained through it.  Therefore the (stationary) system is not in equilibrium.  In fact there is a constant production $\sigma$ of entropy in the environment, in the sense that energy is dissipated in the two thermal baths.  We apply the usual formula \[
 \sigma = J_1/T_1 + J_2/T_2
 \]
  with $J_i$ the energy flux into the $i-$th bath at temperature $T_i$.  Stationarity (conservation of energy) implies that $J_1 + J_2 =0$ so that we can find the direction of the energy current $J_1$ by requiring
  \[
  \sigma = J_1(1/T_1 - 1/T_2)\geq 0\quad \text{ (second law)}\]

Similar scenario's can be written for chemical and mechanical baths that frustrate the system.  Those are typical examples where, to find the direction of the current, we can apply the second law stating that the stationary entropy production be positive\footnote{In the presence of multiple currents we can only require that the matrix of Onsager response coefficients is positive.}.\\
 It is however not uncommon in nonequilibrium to find a system where the direction of the current is essentially not decided by the entropy production.  The paper \cite{mmcs} treats different examples of that.\\

 A simple scenario is to imagine a network consisting of various nodes (vertices) representing each a certain chemical-mechanical configuration of an ensemble of molecules or the coarse-grained positions of diffusing colloids.  The edges between the nodes indicate the possible transitions (jumps) in a continuous time Markov dynamics.  There could be various cycles in that network and some of them relate to the observed or interesting physical or chemical current. 
 The question is what typically will be the direction of that current in that cycle; is it e.g. clockwise or counter-clockwise in the network, which could imply a different direction of the current in physical space.\\
  To be more specific let us look at Fig.~\ref{neckto}, where we see an example of a necklace, a periodic repetition of pearls.  Think of a random walker jumping between the nodes connected via an edge.  Let us suppose we are interested in the current going in the bulk necklace (the red nodes).  The problem becomes non-trivial at the moment we organize the driving in each pearl in such a way that ``entropically'' there is no preferred direction.
  \begin{figure}[h]
   \centering
   \includegraphics[width= 16 cm]{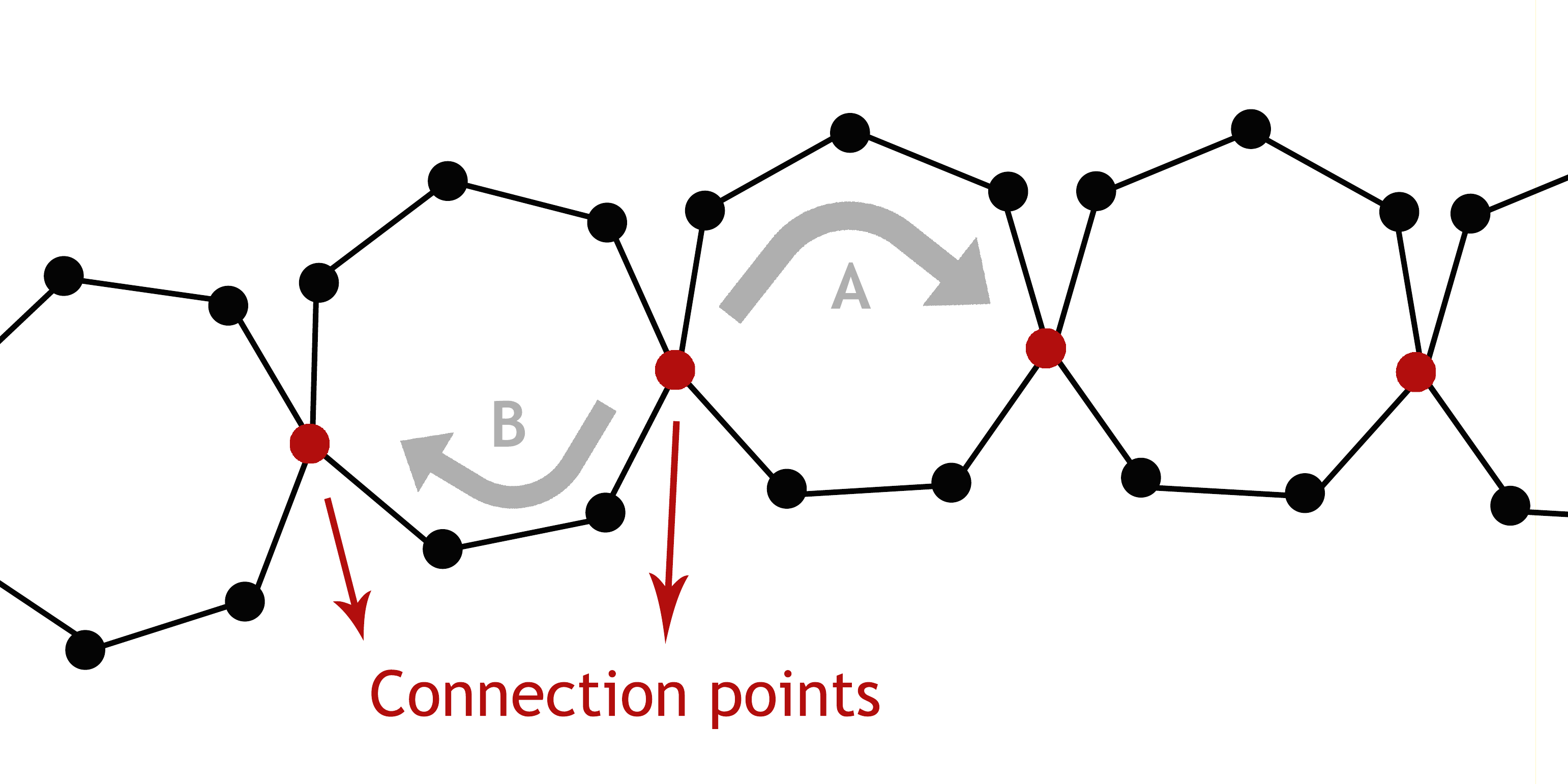}
   \caption{A necklace consists of pearls, here heptagons connected periodically at the red nodes. A clockwise current is generated in each pearl in such a way that the upper $A$-trajectory has the same dissipation as the lower $B$-trajectory.  Nevertheless a current typically appears in the necklace. Courtesy of Mathias Stichelbaut.}
   \label{neckto}
   \end{figure}
   
The simplest example is represented in Fig.~\ref{neck} where the pearls are triangles.  We specify the transition rates as follows:
\begin{eqnarray}\label{arr}
k_{\nearrow} = e^{\varepsilon/2},\qquad k_{\searrow} = e^{\varepsilon/4},\qquad k_\leftarrow = \varphi\,e^{\varepsilon/2}\nonumber\\
k_{\swarrow} = e^{-\varepsilon/2},\qquad k_{\nwarrow} = e^{-\varepsilon/4},\qquad k_\rightarrow = \varphi\,e^{-\varepsilon/2}
\end{eqnarray}
where the arrows reflect the direction of the hopping in each triangle of Fig.~\ref{neck}.   The $\varphi$ is an activity parameter.  
 \begin{figure}[h]
 \centering
 \includegraphics[width= 16 cm]{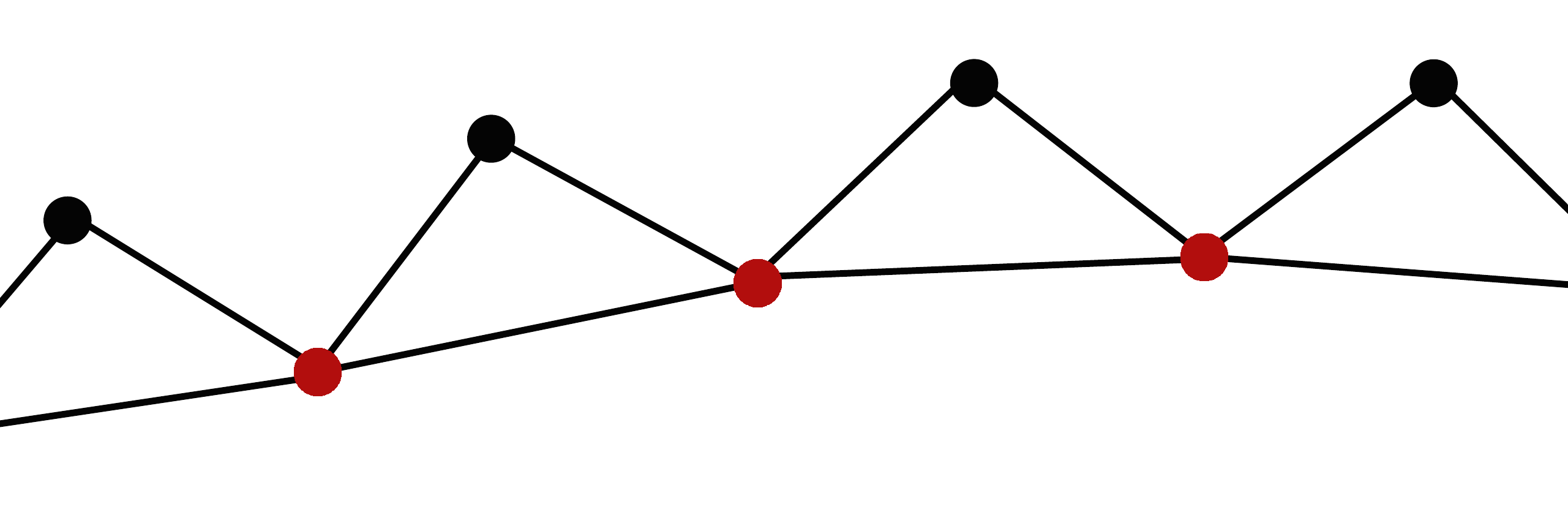}
 \caption{Zooming in on the triangular pearls making the closed necklace.  Will the current go left or right over the lower nodes when an emf with driving $\varepsilon$ in \eqref{arr} is created in each triangle which is entropically neutral? Courtesy of Mathias Stichelbaut.}
 \label{neck}
 \end{figure}
 We have chosen the rates in such a way that the trajectory $A = \nearrow \searrow$ has entropy flux
 \[
S(A)= \log\frac{k_\nearrow \,k_\searrow}{k_\nwarrow \,k_\swarrow} = \varepsilon
 \]
 identical to the entropy flux over trajectory $B = \leftarrow$, $S(B) = \log\frac{k\leftarrow}{k_\rightarrow} =\varepsilon$.
 
 Many other choices are possible of course to achieve that.  Note that the entropy flux balance between A and B is independent of $\varphi$ which is a time-symmetric parameter in the sense that $k_\leftarrow k_\rightarrow =\varphi^2$.  A non-dissipative effect would be to see how changes in $\varphi$ influence the nonequilibrium nature of the system, here in particular, how it can decide the direction of the necklace current.  And in fact it does: for a fixed driving $\varepsilon$, by changing $\varphi$ we can change the direction of the current.  As an example, we look at Fig.\ref{j3}: we see the necklace current $J_{\rightarrow}$ as function of the driving $\varepsilon$ for two different values of $\varphi$.  Not only is the current not monotone in $\varepsilon $ for $\varphi>1/2$, it also changes sign with different stalling points appearing for different $\varphi$.  With $\kappa = \exp(\varepsilon/4)$, the necklace current equals
  \begin{equation}
  J_{\rightarrow}(\varphi,\varepsilon)= \displaystyle{ (\kappa^4 - 1)(1 - \varphi(1+\kappa^{-1}))  \over \kappa^3 + \kappa^2 + \kappa + 1}
  \label{eqMC:currtriangle}
  \end{equation}
 Note that  there is stalling, $J_{\rightarrow}=0$, when $\varphi = \kappa/(1+\kappa)$ (requiring  $\varphi \in (1/2,1)$ for $\varepsilon>0$).  All the same the stationary distribution does not depend here on $\varphi$.\\
  The same and more complicated changes happen for more complicated necklaces but the example with triangles makes clear that  non-dissipative activity-parameters such as $\varphi$ can play a crucial role. 
 
\begin{figure}[h]
 \centering
 \includegraphics[width= 10 cm]{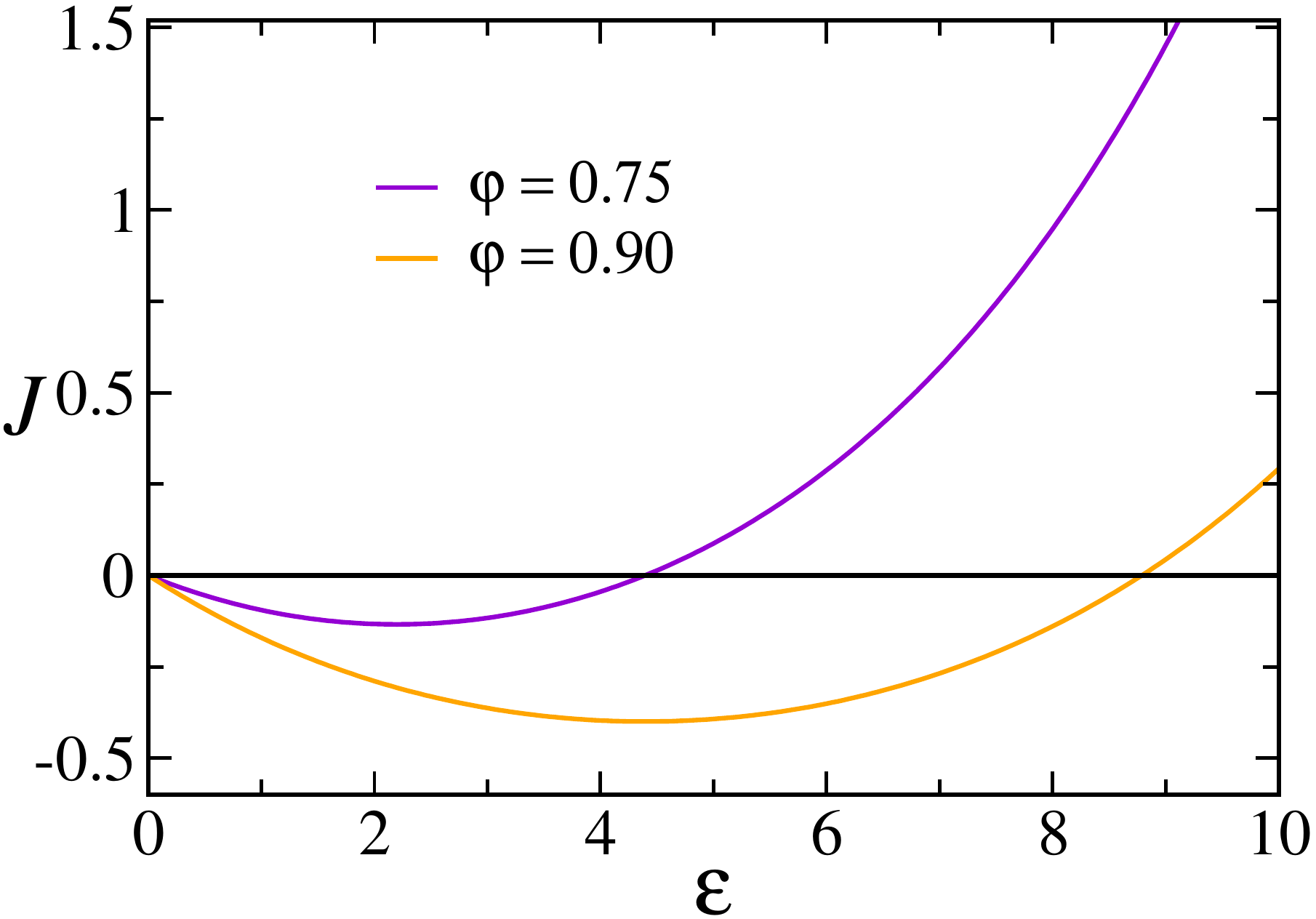}
 \caption{The current towards the right for the triangular necklace in Fig.~\ref{neck}.  Courtesy of Urna Basu.}
 \label{j3}
 \end{figure}

 We could have  also more cycles in the system, and depending on the cycle  a particular current would go one way or the other, and yet both directions show exactly the same entropy production.  We give a more complex but also biophysically more relevant illustration of the same phenomenon, {\it cf.} also \cite{physA}.\\
   Consider Fig.~\ref{fig1}; it refers to a simple model for the stepping of a molecular motor, here Myosin.  Myosin V is a very well studied processive cargo transporter with two heads acting as feet for head--over-head movement along some track, and playing a role in muscle contraction.  Its fuel is ATP and Fig.~\ref{fig1} describes the power stroke.  There is no need to enter here the biophysical and biomechanical details, but the issue is that {\it a priori} it is of course not so that the ATP consumption is a driving similar to installing a gradient in Fig.\ref{hecs}.  The question is again, what determines the direction of the current, and the answer is (again) that there is an important non-dissipative ingredient.  Let us make abstraction of the chemistry and concentrate on a simplified model.\\
  Suppose a Markov jump process with six states $K=\{D,x,v,T,w,y\}$; see Fig.~\ref{fig1}(a) reproduced from \cite{physA}.
The rates for the transitions $D\rightarrow v \rightarrow T\rightarrow y\rightarrow D$ are taken to be
\[
k(D,v) = a,\quad k(v,T) = \psi_1,\quad k(T,y) = \psi_2\,e^{s_2},\quad k(y,D) = d\,e^{s_3}
\]
\[
k(v,D) = a\,e^{s_0},\quad k(T,v) = \psi_1 \,e^{-s_1},\quad k(y,T) = \psi_2,\quad k(D,y) = d
\]
The $s_0, s_1, s_2, s_3$  are entropy fluxes  (always per $k_B$) over the corresponding jumps. They are thermodynamically decided by the reactions involving the different chemical potentials of the various substances or molecules plus some extra chemical driving to make it a nonequilibrium system. Similarly, for transitions $D\rightarrow w \rightarrow T\rightarrow x\rightarrow D$, we have
\[
k(D,w) = b,\quad  k(w,T) = \psi_1,\quad k(T,x) = \psi_2 \,e^{s_2}\quad k(x,D) = c\,e^{s_3}
\]
\[
k(w,D) = b\,e^{s_0},\quad  k(T,w)= \psi_1\,e^{-s_1},\quad k(x,T) = \psi_2,\quad k(D,x) = c
\]
The numbers $a,b,c,d,\psi_1,\psi_2$ are the activity parameters, an essential ingredient in the reactivities.  We have also included some symmetry at least concerning dissipative aspects of the transitions; over the dotted line in Fig.~\ref{fig1}: for example $k(w,T)/k(T,w) = k(v,T)/k(T,v) = \exp s_1$.

Look now at the two trajectories in Fig.~\ref{fig1}(a) touching each other at the states $D$ and $T$.  They are $R_1=(D,v,T,y,D)$ and $R_2= (D,w,T,x,D)$, each other's reflection over the dotted horizontal line in Fig.~\ref{fig1}(a). 

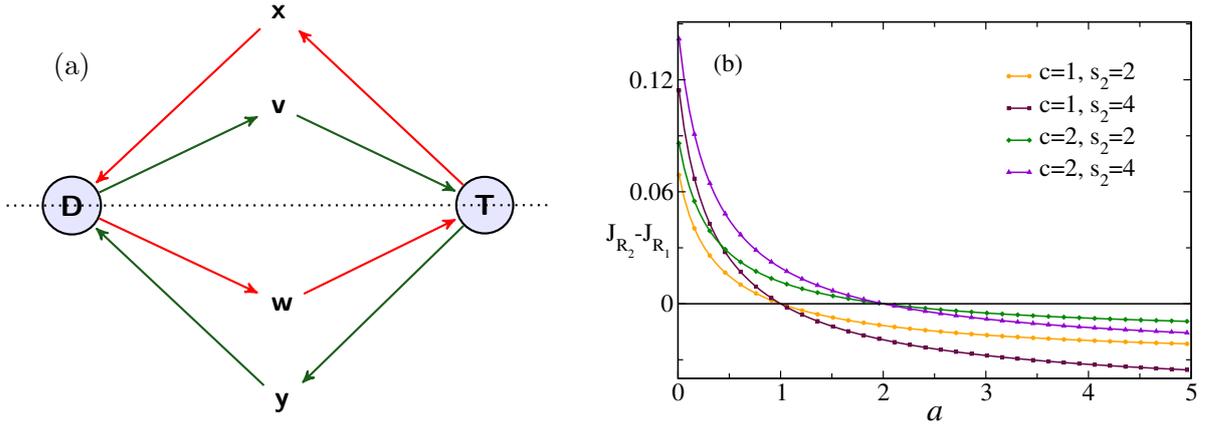
\begin{figure}[h!]
\centering
 \begin{tikzpicture}
[-,>=stealth',shorten >=1pt,auto,node distance=1cm,
   thick,main node/.style={circle,fill=blue!30,draw,font=\sffamily\small\bfseries},main node/.style={circle,fill=blue!10,draw,font=\sffamily\small\bfseries},state/.style={font=\sffamily\small\bfseries}]
  \node[state] (1) {v};
  \node[main node] (2) [below left=0.8 cm and 2.2 cm of 1] {D};
  \node[state] (3) [below right=0.8 cm and 2.2 cm of 2] {w};
  \node[main node] (4) [below right=0.8 cm and 2.2 cm of 1] {T};
  \node[state] (5) [above= 0.8 cm of 1] {x};
  \node[state] (6) [below =0.8 cm of 3] {y};
    \node[] (7) [left of=2 ] {};
     \node[] (8) [right of=4]{};
   \node[] (10) [above=1.1 cm of 2]{(a)};
    \path [dotted] (7) edge  (8);
  \path[every node/.style={font=\sffamily\small}]
     (2) edge  [green!30!black!90,->] node {} (1)
    (1)   edge [green!30!black!90,->] node {} (4)
     (4)   edge [green!30!black!90,-> ]node {} (6)
      (6)  edge [green!30!black!90,-> ]node {} (2)
  (3) edge [red,->](4)
  (2) edge[red,->] (3)
  (5) edge [red,->](2)
  (4) edge [red,->] (5);
\end{tikzpicture}
\hspace*{0.2 cm}
 \includegraphics[width= 7.8 cm]{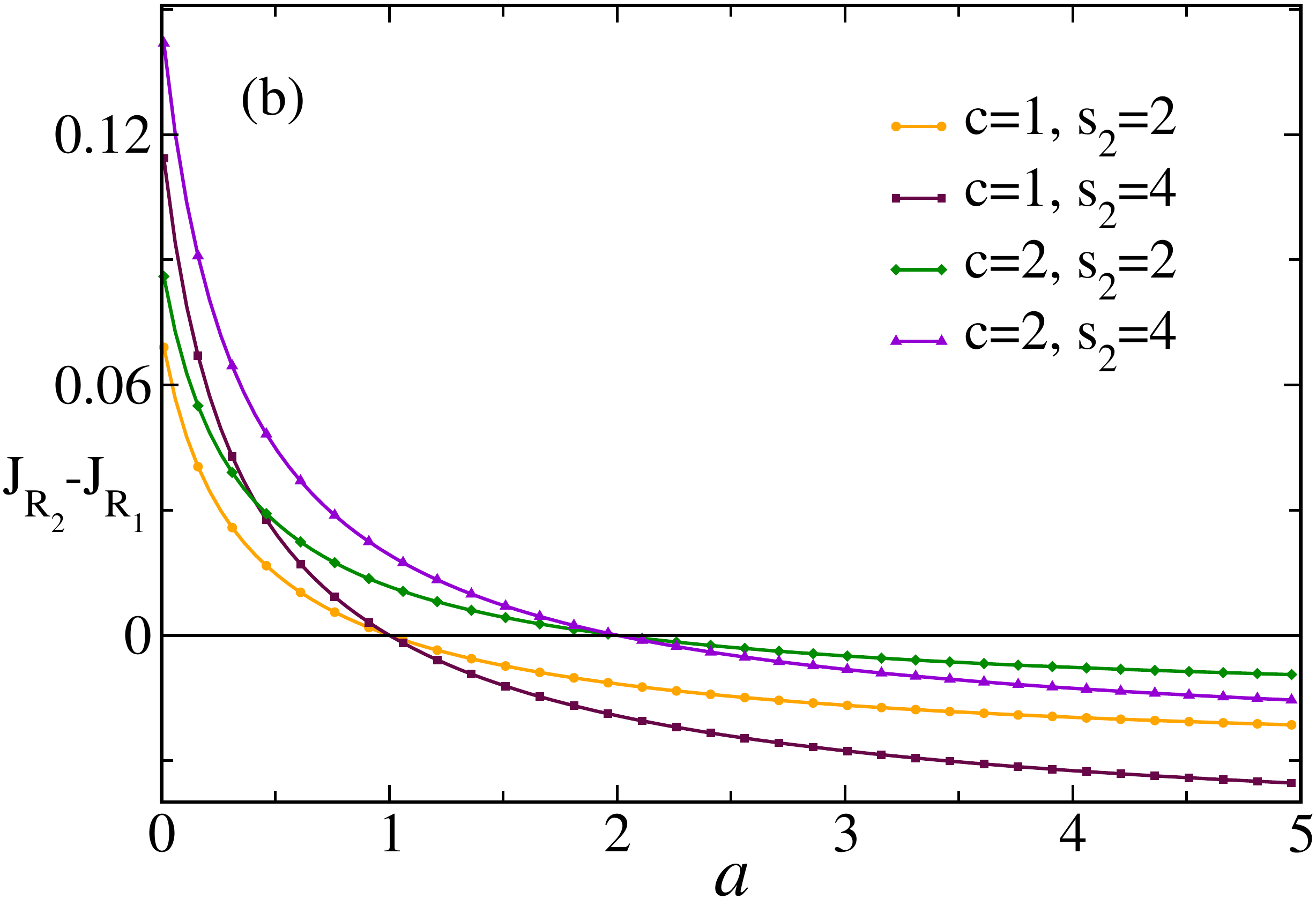}
\caption{(a) Reproduction of Fig.~3 in \cite{physA}. Cycles over $R_1$ (clockwise) and $R_2$ (counter-clockwise). (b) The parameters are $\psi_{1,2}=1, b=c, s_0 = s_1 = 1$ and $s_3= -1$. The vertical axis shows the counter-clockwise minus the clockwise current as a function of $a=d$. }
\label{fig1}
\end{figure}

Note that the entropy fluxes $S_1$ and $S_2$ respectively for going through $R_1$ and for going through $R_2$, are exactly equal:
\begin{equation}\label{0123}
S_1 = -s_0 + s_1 + s_2 +s_3 = S_2
\end{equation}
 and similarly, for the reversed cycles in each ring. As a consequence, the sign of the entropy flux does not determine here the direction of the current: $S_1>0$ whenever $S_2>0$ but the corresponding currents cycle in opposite directions.  Just from the perspective of dissipation, as a positive entropy flux can be equally realized by clockwise  or by counter-clockwise  turning, we cannot conclude the direction of the current.  What truly decides here the direction is not-entropic; it is frenetic, see Fig.~\ref{fig1}(b).  We see how the direction of the current is entirely decided by the prefactors, which are symmetric: for $a=d>b=c$ the system cycles over $R_1$, and for $a=d<b=c$ it is instead cycle $R_2$ which is preferred.  The asymmetry between the two cycles resides in the prefactors of the reaction rates, e.g. $a\neq b$ for the transition  $D\rightarrow v$ {\it versus} $D\rightarrow w$.  When there is detailed balance, i.e. for  $s_0=s_1+s_2+s_3$, the discrepancy $a\gg b$ or $c\neq d$ etc.  is irrelevant and the stationary regime would be equilibrium-dead showing no orbiting whatsoever.
Those activity considerations for the nonequilibrium statistical mechanical aspects of Myosin V motion are discussed in \cite{physA}.
General considerations on how dynamical activity plays in determining the direction of ratchet currents are found in \cite{woj}.

\subsection{Negative differential conductivity}

The usual way random walks are discussed is by giving the rates for the walker to move to its neighbors.  Let us then take the simple 1-dimensional walk, $x\in \bbZ$,  with rates $k(x,x+1)=p$ and $k(x,x-1)=q$.  The fact that $p>q$ would mean that there is a bias to move to the right.  But suppose we now have a real motion of quasi-independent particles moving a in tube with some obstacles, much like in Fig.~\ref{traj}.

\begin{figure}[h]
\centering
\includegraphics[width= 16 cm]{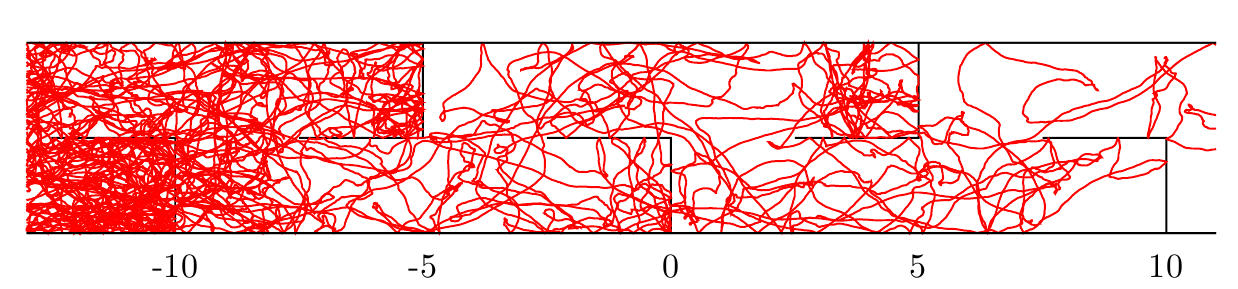}
\caption{Trajectories of driven particles, more and more trapped in the obstacles as the driving gets bigger. Courtesy of Soghra Safaverdi.}
\label{traj}
\end{figure}

We see there the trajectories of particles being pushed to the right, while Joule heating the environment.
We imagine periodic boundary conditions so that the constant external field is rotational (and hence cannot be derived from a potential, which causes the breaking of detailed balance).
There are hooks on the floor and at the ceiling of the tube what are obstacles for the particles' motion.
Suppose we want to relate that situation to a random walk; what would we choose for $p$ and $q$?
The external field delivers a certain work $W$ per unit time to each particle, which is dissipated in the environment at inverse temperature $\beta$.  Locally, we can take the system to obey detailed balance \eqref{ldb} and thus we require
\[
\frac{p}{q} = e^{\beta W}
\] 
That does not quite determine the rates $p,q$.  What appears important here also is to know how the escape rates depend on $W$; that is the dependence
\[
p+q = \psi(W)
\]
From the situation in Fig.~\ref{traj} it is clear that a large driving field or what amounts to the same thing, large $W$, causes trapping.  The particles are being pushed against the hooks, or caught and have much difficulty in escaping from the trap as they are constantly pushed back on them. Hence the escape rates and the function $\psi$ will be decreasing in $W$.  In fact, we can expect that here $\psi(W) \sim \exp{-\beta W}$; see \cite{pieter}.\\

Let us see what that means for the current; the random walker has a current
\begin{equation}\label{jj}
J= p-q = \psi(W)\, \frac{1-e^{-\beta W}}{1+e^{-\beta W}}  
\end{equation}
We clearly see that for large $W$, in fact basically outside the linear regime in $W$, the current becomes decreasing in $W$ (whenever $\psi$ is decreasing in $W$); the more you push the more the particles get trapped and the current goes down; see also \cite{zia}.\\

The above mechanism, described here in the simplest possible terms, is an instance of negative differential conductivity, the current showing a negative derivative with respect to the driving field. It is physically interesting and also very important, but for us it suffices to remark that the effect is clearly non-dissipative.  Of course one needs a current in the first place but what happens with it is related to time-symmetric fluctuations, here in the form of waiting time distributions.  The dynamical activity is a function of $W$ and cannot be ignored, and is even essential in the understanding of the effect of negative differential conductivities.  We have a general response--approach to that effect coming up in Section \ref{reps}; see \eqref{6n}--\eqref{cmu}. \\

As a final remark, the trapping effect in the above is trivially caused by the geometry.  There are other mechanisms like in the Lorentz gas where the obstacles are random placed or even moving slightly, \cite{urna,sarrac,bsv}.  But we can also imagine that that trapping and caging is not caused by external placements but by the interaction itself between the driven particles.  In that case we speak about jamming and glass transitions \cite{chan}, or even about many body localization \cite{gar}.

\subsection{Death and resurrection of a current}
The previous section can be continued to the extreme situation where the current actually dies (vanishes completely) at large but finite driving.  That phenomenon has been described in the probability and physics theory alike; see e.g. \cite{db,sol}.  Recently we showed how the current can resurrect for those large field amplitudes by adding activity in the form of a time-dependent external field, \cite{thi}.  The cure is somewhat similar to a dynamical counterpart of stochastic resonance, \cite{scholar}.  The point is that by the `shaking' of the field, particles get again time to go backwards and find new ways to escape.  It is effectively a resetting mechanism that improves the finding of the exit from the condition which previously meant an obstacle.\\

One should imagine an infinite sequence of potential barriers (along a one-dimensional line). There is an external field $E$ which drives the particles to the right but at the same time the height of the potential barriers grows with $E$.   In other words we have again an activity parameter (for escape over the barrier) here which decreases with $E$.  Since there is an infinite sequence of them the current to the right will vanish identically when the field $E$ reaches a threshold $E_c$.  But suppose we now add a time-dependence and make the external field $\cal E = E f(t)$ where $f(t)$ is periodic and $\int_0^\tau f(t)/\tau = 1$ over the period $\tau$.  The amplitude of the force over the period has therefore not changed.  Yet, we can increase the variance of $f(t)$ for fixed such amplitude, and hence the capacity of negative field will grow and the potential barrier goes down, even compensating for the negative field, so that the particle can diffuse through to the right.  The result of that activation (the ``shaking'') is the resurrection of the current in the form of a first order phase transition (at zero temperature); see \cite{thi}.

\section{Response}\label{reps}

We come to the meaning and the extension of the fluctuation--dissipation theorem. The general aim is the physical understanding of the statistical reaction of a system to an external stimulus.  In that sense we look at averages of certain quantities, for example over many repeated measurements.  We start from a stationary condition and we perturb the system over some time period.  The question is to find a good representation of the response of the system: we want to learn what determines the susceptibility in terms of the original (unperturbed) system.  Old examples are the Sutherland-Einstein relation between diffusion and mobility, or the Nyquist-Johnson formula for the thermal noise in resistors.  These can be extended to nonequilibrium situations and invariantly specific non-dissipative effects show up.  Alternatively, measuring response can inform us about activity parameters and time-symmetric traffic in the original system.\\

For the above purpose and especially for nonequilibrium systems, working on space-time is more convenient than directly working on the single-time distribution.  The path-space distribution is local in space-time and has often explicit representations; we speak then about dynamical ensembles and it is part of nonequilibrium statistical mechanics to learn how they are specified and what they determine.

\subsection{Standard fluctuation--dissipation relation}\label{path}

We start by describing the path-space approach for characterizing the response in equilibrium.  We here thus deviate from the usual rather formal analytic treatment which is on the level of perturbations of the time-flow, e.g. via a Dyson formula for the perturbation of semi-groups or unitary evolutions.\\
A dynamical ensemble gives the weight of a trajectory of system variables. Let $\omega$ denote such a system trajectory over time-interval $[0,t]$.  It could be the sequence of positions of an overdamped colloid or the sequence of chemical or electronic states of a complex molecule etc.
We consider a path--observable $O=O(\omega)$ with expectation 
\begin{equation}\label{sta}
\langle O \rangle = \int {\mathbb D}[\omega] \, P(\omega)\,O(\omega) =  \int {\mathbb D}[\omega] \,e^{-\cal A(\omega)}\,P_{\text{eq}}(\omega)\,O(\omega)
\end{equation}
Here ${\mathbb D}[\omega]$ is the notation, quite formally, for the volume element on path--space.
The perturbed dynamical ensemble is denoted by $P$ and gets specified by an action $\cal A$ with respect to the reference equilibrium ensemble $P_{\text{eq}}$:
\[
P(\omega) = e^{-\cal A(\omega)}\,P_{\text{eq}}(\omega)
\]
At the initial time, say $t=0$, the system is in equilibrium and the path-probability distributions $P$ and $P_{\text{eq}}$ differ (only) because $P$ is the dynamically perturbed ensemble.  Time-reversibility of the equilibrium condition is the invariance $P_{\text{eq}}(\theta\omega) = P_{\text{eq}}(\omega)$ under time-reversal $\theta$, defined on paths $\omega = (x_s, 0\leq s\leq t)$ via
\[
(\theta\omega)_s =\pi x_{t-s}
\]
with kinematical time-reversal $\pi$ (e.g. flipping the sign of velocities) as in Fig.~\ref{kine}.\\

 \begin{figure}[h]
 \centering
 \includegraphics[width= 12 cm]{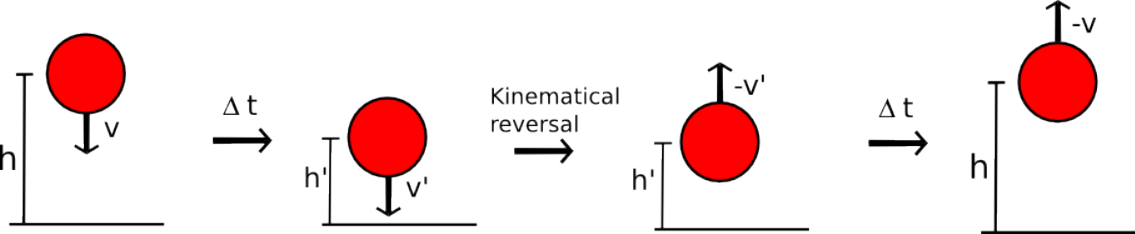}
 \caption{Time-reversal in free fall.}\label{kine}
 \end{figure}

We decompose the action $\cal A$ into a time-antisymmetric $S(\omega)$ and a time-symmetric $D(\omega)$ part,  
\begin{equation}\label{deco}
 \cal A = D-S/2
 \end{equation}
  with $S\theta=-S$ and $D\theta=D$.  These are the  entropic ($S$) and  frenetic ($D$) components of the action and they respectively give excesses in entropy fluxes and in dynamical activity as caused by the perturbation. In that way $S(\omega)$ and $D(\omega)$ depend on time $t$ because they are defined on paths $\omega$ in the time-interval $[0,t]$.  For the present linear response around equilibrium, all will be decided by $S$ (as we will see below). The reason why that $S$ is related to the entropy flux amounts to \eqref{ldb} and the understanding of the origin of the local detailed balance condition; see e.g. \cite{poincare}.  For example, in case of a potential perturbation of strength $\varepsilon$, where we change the energies $E(x)\rightarrow E(x) - \varepsilon V(x)$, we would have $S(\omega) = \varepsilon\beta \,[V(x_t) - V(x_0)]$. In case of an external field $\varepsilon$ which induces a single current $J$, we would have $S(\omega)= \varepsilon\,J(\omega)$. In the case the trajectory is very special, where $\omega=\theta \omega$, we have ${cal A}(\omega) = D(\omega)$ and the path-probability is given from the frenetic component only. It contains the waiting times (when the system sojourns in some state) and the activity parameters.\\

Expanding \eqref{sta} to linear order in the perturbation around equilibrium yields
\begin{equation}\label{2n}
\langle O \rangle = \langle O \rangle_{\text{eq}} - \langle {\cal A}(\omega)\, O(\omega)
\rangle_{\text{eq}}
\end{equation}
in terms of the equilibrium expectations
\[
\langle g(\omega)\rangle_{\text{eq}} :=  \int {\mathbb D}[\omega] \,P_{\text{eq}}(\omega)\,g(\omega)
\]
For the time-reversed observable that gives
\begin{equation}\label{2n2}
\langle O \theta\rangle = \langle O \rangle_{\text{eq}} - \langle {\cal A}(\theta\omega)\, O(\omega)
\rangle_{\text{eq}}
\end{equation}
where we have used the time-reversibility $P_{\text{eq}}(\theta\omega) = P_{\text{eq}}(\omega)$ so that $\langle g(\omega)\rangle_{\text{eq}} = \langle g(\theta\omega)\rangle_{\text{eq}} $.
Subtracting \eqref{2n2} from \eqref{2n} gives
\[ 
\langle O - O\theta\rangle = -\langle [{\cal A}(\omega)-{\cal A}(\theta\omega)]\, O(\omega)
\rangle_{\text{eq}} 
\] 
always to linear order in the perturbation.
Now use that the time-anisymmetric part ${\cal A}(\omega)-{\cal A}(\theta\omega) = - S(\omega)$ equals the entropy flux, to get
\begin{equation}\label{2n4}
\langle O - O\theta\rangle = \langle S(\omega)\, O(\omega)
\rangle_{\text{eq}}
\end{equation}
for all path-variables $O$.\\ 
For an observable $O(\omega)=O(x_t)$ that depends on the final time, we have $O\theta(\omega)= O(\pi x_0)$.  Since $\langle O(\pi x_0)\rangle_{\text{eq}} = \langle O(x_0)\rangle_{\text{eq}} = \langle O(x_t)\rangle_{\text{eq}}$  we receive
\begin{equation}\label{kubo2}
\langle O(x_t)\rangle - \langle O(x_t)\rangle_{\text{eq}} =  \langle S(\omega)\,O(x_t)
\rangle_{\text{eq}} 
\end{equation}
In other words, the response is completely given by the correlation with the dissipative part in the action, the entropy flux $S$.  That result \eqref{kubo2} is the Kubo formula even though it is often presented in a more explicit way.  Suppose indeed that the entropy flux is 
$S(\omega) = \varepsilon\beta \,[V(x_t) - V(x_0)]$ (tiem-independent perturbation by a potential $V$ starting at the initial time zero) as mentioned above formula \eqref{2n}, then we would get from \eqref{kubo2} that
\[
\langle O(x_t)\rangle = \langle O(x_t)\rangle_{\text{eq}} +  \varepsilon\beta \,\langle [V(x_t) - V(x_0)]\,O(x_t)
\rangle_{\text{eq}} + \,\,\text{ terms of order  } \varepsilon^2 
\]
Similarly, for such a time-dependent (amplitude $h_s$) perturbation that becomes
\begin{equation}\label{kuboa}
\langle O(x_t)\rangle = \langle O(x_t)\rangle_{\text{eq}} +  \varepsilon\beta \,\int_0^t\id s \,h_s\,\frac{\id}{\id s}\langle V(x_s)\,O(x_t)
\rangle_{\text{eq}} \,+ \,\text{ terms of order  } \varepsilon^2 
\end{equation}
When the observable is odd under time-reversal, $O(\theta\omega) = -O(\omega)$ like $O(\omega)=S(\omega)$ the entropy flux itself, or  when $O(\omega) = J(\omega)$ is some time-integrated particle or energy current, then we get from \eqref{2n4}  the Green--Kubo formula
\begin{equation}\label{gk}
\langle J\rangle = \frac{1}{2} \,\langle S(\omega)\, J(\omega)
\rangle_{\text{eq}}, \quad \text{ and } \;\; \langle S\rangle = \frac{1}{2} \,\text{ Var}_{\text{eq}}[S]
\end{equation}
That the linear order gets expressed as a correlation between the observable in question and the entropic component $S$ only is the very essence of the fluctuation--dissipation relation. One can imagine different perturbations with the same $S(\omega)$ (dissipatively equivalent) and there will the same linear response around equilibrium. Obviously, in the time-correlation functions that enter the response, there are kinetic aspects and non-dissipative contributions are thus present already in first order.  Yet, there is no explicit presence of non-dissipative observables.  Again, even when the perturbation depends on kinetic and time-symmetric factors, the linear response \eqref{kubo2})--\eqref{kuboa}--\eqref{gk} erases that detailed dependence and only picks up the thermodynamic dissipative part $S$.  In particular, in equilibrium for the Gibbs distributions the stationary distribution is just energy-entropy determined.

\subsection{Enters dynamical activity}

The previous section discussed the linear response around equilibrium (by using its time-reversibility).  Yet, the line of reasoning is essentially unchanged when doing linear response around nonequilibrium regimes.  That is, up to equation \eqref{2n} nothing changes:
\begin{equation}\label{4n}
\langle O \rangle = \langle O \rangle_{\text{ref}} - \langle {\cal A}(\omega)\, O(\omega)
\rangle_{\text{ref}}
\end{equation}
where the subscript ``ref'' in the right-hand side expectation simply replaces the equilibrium ones.  That new reference is for example steady nonequilibrium where (in the left-hand side) we investigate the response; see \cite{fdr} for more details.\\

We still do the decomposition \eqref{deco} where the $S$ respectively the $D$ now refer to excesses in entropy flux and in dynamical activity with respect to the unperturbed nonequilibrium steady condition.
Substituting that into \eqref{4n} we simply get
\begin{equation}\label{6n}
\langle O \rangle - \langle O \rangle_{\text{ref}}= \frac 1{2}\langle S(\omega)\, O(\omega)
\rangle_{\text{ref}} - \langle D(\omega)\, O(\omega)
\rangle_{\text{ref}}
\end{equation}
and a frenetic contribution with $D=D\theta$ enters as second term in the linear response.  That is a non-dissipative term as it involves time-symmetric changes, in particular related to dynamical activity and time-symmetric currents.  Remark also that for no matter what initial/reference distribution
\begin{equation}\label{9n}
 \frac 1{2}\langle S(\omega)
\rangle_{\text{ref}} = \langle D(\omega)
\rangle_{\text{ref}}
\end{equation}
by taking $O\equiv 1$ in \eqref{6n}.  That constitutes a useful calibration for measurements.
\\

If the observable is a current $J$ which is caused by a constant external field $W$, and we change $W\rightarrow W+ \id W$ as the perturbation in the reference process, so that $S(\omega) =\beta \id W \,J(\omega)$, then \eqref{6n} gives
\begin{equation}\label{cmu}
\frac{\id \langle J\rangle}{\id W} = \frac{\beta}{2}\langle J^2(\omega)\rangle_{\text{ref}} - \langle D(\omega)\, J(\omega)
\rangle_{\text{ref}}
\end{equation}
which is the modification of the Sutherland--Einstein relation under which mobility (left-hand side) is no longer proporitonal to the diffusion constant (first term on the right-hand side), cf. \cite{roy1,roy2}.  The correction (second term in right-hand side) is a non-dissipative effect in the correlation between current and dynamical activity. As an important example it shows that negative differential conductivity can only be the result of a (large) positive correlation in the unperturbed system between the current and the excess dynamical activity.  That requires breaking of time-reversal invariance surely, otherwise $\langle D(\omega)\, J(\omega)
\rangle_{\text{ref}}=0$.  For the simple random walk under \eqref{jj} the excess dynamical activity is essentially minus the change in escape rate $\psi'(W)$ times the number of jumps.  If that change in escape rate is sufficiently negative, we get negative differential conductivity, because the number of jumps correlates positively with the current.

\subsection{Second order response}

We consider next the extension to second order of the traditional Kubo formula for response around equilibrium, \cite{pccp,beh}.  For order-bookkeeping we suppose that the perturbation or external stimulus is of strength $\varepsilon \ll 1$ and is present in the action ${\cal A} = {\cal A}_\varepsilon$, depending smoothly on  $\varepsilon$. We restrict us here also to the case where the perturbation is time-independent (after the initial time).\\
Furthermore we assume that the perturbation enters  at most linearly in $S$, i.e., higher derivatives like $S''_{\varepsilon=0} =0$ are zero; it means that the Hamiltonian is linear in the perturbing field. We refer to \cite{pccp} for the details. 

The result that extends \eqref{kubo2}
is
\begin{equation}\label{kubo3}
\langle O(x_t)\rangle - \langle O(x_t)\rangle_{\text{eq}} =  \varepsilon\,\langle S'_0(\omega)\,O(x_t)
\rangle_{\text{eq}} - \varepsilon^2\,\langle D'_0(\omega) \,S'_0(\omega)\, O(x_t)\rangle_{\text{eq}}\Oe.
\end{equation}
where the primes refer to $\varepsilon-$derivatives at $\varepsilon=0$.  We see that at second order around equilibrium the excess dynamical activity $D'_0(\omega) $ enters the response, and will of course have its consequences as non-dissipative effect.\\

For an example we take the zero range model, representing  a granular gas in one dimension; see \cite{harrisevans}.
\begin{figure}[t]
 \centering
\includegraphics[width=12 cm]{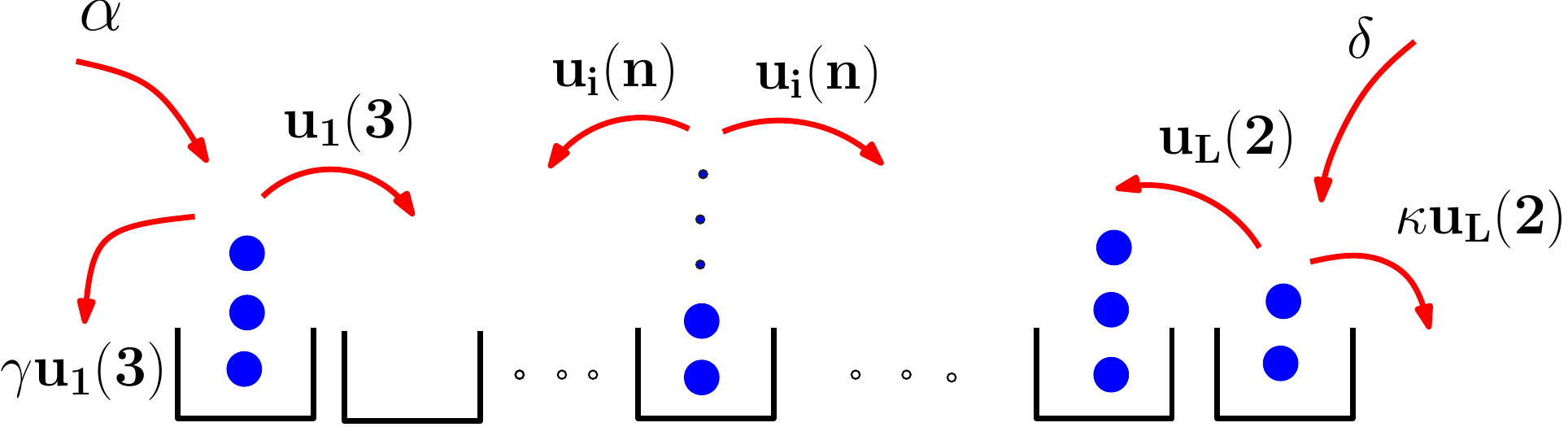}
 \caption{ Schematic representation of the open zero range process with respective transition rates as they depend on the local particle number $n$.  The chemical potential $\mu$ of the environment is given in \eqref{chemp} but does not fully determine the entry and exit rates.}
 \label{zrp}
\end{figure}
There are $L$ sites and each is occupied with $n_i$ number of particles, see Fig.~\ref{zrp}.  In the bulk a particle hops from site $i$ to its right or left neighbor at rate $u(n_i)$.  The boundary sites are connected with particle reservoirs, their action being modeled via entrance and exit rates at the left and right boundary sites $i=1$ and $i=L$.  Entry rates are $\alpha$ and $\delta$ respectively; exit rates are $\gamma\, u(n_1)$ from the first site and $\kappa\, u(n_L)$ at the last site.  We have global detailed balance and no stationary current when $\alpha\kappa= \gamma \delta$.  Then, the environment is thermodynamically specified by the chemical potential $\mu$ and inverse temperature $\beta$ via
\begin{equation}\label{chemp}
\alpha/\gamma =  \delta/\kappa = e^{\beta \mu}
\end{equation}
Yet, just as for our random walker above, the jump rates $\alpha,\delta,\gamma$ and $\kappa$ are not completely specified by $\beta\,\mu$; the coupling with the environment indeed depends on more kinetic elements.\\

We want to investigate the response to a change in chemical potential $\mu\rightarrow \mu+ \varepsilon $ of the bath.
The dissipative or entropic component to the response is completely specified, no matter {\it how} that change is implemented via the entry and exit rates.  We always have $S_0^\prime = \beta J_{\text{in}}$ in terms of   the net number  $J_{\text{in}}$  of particles transferred into the system from the environment during the time interval $[0,t]$.  However, starting from second order, the non-dissipative component enters and we must see for $D_0^\prime$.\\  A first possibility is that only the entry rates at both edges are changing $\alpha \to \tilde \alpha = \alpha e^{\beta \varepsilon}, \delta \to \tilde \delta = \delta e^{\beta \varepsilon}$ for an increase in the chemical potential of $\tilde{\mu}=\mu + \varepsilon.$
Then, we can calculate
\begin{equation}\label{34}
D_0^\prime = \beta (\alpha+\delta) t -\frac \beta 2 \,{\cal I}
\end{equation}
Here ${\cal I}$ is the total number of particle exchanges between the system and the reservoir during $[0,t]$, again non-dissipatively, time-symmetric.  We also see the explicit dependence on the entry rate $\alpha+\delta$.
The second order response is obtained by inserting \eqref{34} into \eqref{kubo3},
\begin{equation}
\langle S_0'(\omega ) D_0'(\omega ) O(t) \rangle_{\text{eq}} = \beta^2 \left[(\alpha+\delta) t \langle J_{\text{in}} \, O(t)\rangle_{\text{eq}}  - \frac 12 \langle J_{\text{in}}\, {\cal I} \, O(t)\rangle_{\text{eq}} \right]\label{eq:pert1}
\end{equation}
If we consider a different perturbation where both the entry and the exit rates are modified $\alpha \to \tilde \alpha = \alpha e^{\beta \varepsilon/2}/(1+\beta \varepsilon), \gamma \to \tilde \gamma = \gamma e^{-\beta \varepsilon/2}/(1+\beta \varepsilon)$ and $\delta \to \tilde \delta = \delta e^{\beta \varepsilon/2}/(1+\beta \varepsilon), \kappa \to \tilde \kappa = \kappa e^{-\beta \varepsilon/2}/(1+\beta \varepsilon)$, while we still get the same shift in the chemical potential $\tilde{\mu}=\mu + \varepsilon$, the frenetic part now has
\[ 
D_0^\prime(\omega) = \beta  \left[{\cal I}  - \frac 12 (\alpha+\delta) t -\frac 32 \int_0^t \id s~[\gamma u(n_1(s))+ \kappa u(n_L(s))]\right]
\] 
and
\begin{eqnarray*}
\langle S_0'(\omega ) D_0'(\omega ) O(t) \rangle &=& \beta^2 \left[\langle J_{\text{in}}{\cal I};O \rangle_{\text{eq}} - \frac 12 (\alpha+\delta) t \langle J_{\text{in}} ; O(t)  \rangle_{\text{eq}} \right. \cr
&& -\frac 32  \left. \int_0^t \id s~ \langle [\gamma u(n_1(s))+\kappa u(n_L(s))]\,J_{\text{in}}; O(t)\rangle_{\text{eq}}\right] \label{eq:pert2}
\end{eqnarray*}
Here we see the explicit appearance of time-symmetric observables from second order response on.  The linear order response is exactly the same for the two types of perturbations, indistinguishable because that is purely dissipative.\\
Experimental verification and use of the second order response as discussed here was done in \cite{beh}; there the authors show that the non-dissipative contribution is measurable separately.

\subsection{Breaking of local detailed balance}

We have in all examples insisted on the assumption \eqref{ldb} of local detailed balance.  That is the modeling hypothesis so far for nonequilibrium (driven) systems.  Here is however the appropriate place to show the limitations of that assumption.\\
Suppose indeed we consider a probe in contact with a nonequilibrium system as the ones we are discussing in these notes.  The system interacts with the probe via a joint interaction potential $U(q,x)$ appearing in the total Hamiltonian. We denote by $q$ the probe's position and $x$ are the degrees of freedom of the medium. Clearly now the probe will feel friction and noise from the nonequilibrium medium much in the same way as a Brownian particle or a colloid suspended in an equilibrium Newtonian fluid.  The relation between friction and noise for the latter satisfies the Einstein relation, also  called the second fluctuation-dissipation relation.  That in turn is responsible for the probe effective evolution equation verifying local detailed balance as given for discrete processes in \eqref{ldb}.  But that fails altogether for a probe in a nonequilibrium medium.  The relation between friction and noise is no longer the standard Einstein relation then, and the effective probe dynamics will not satisfy local detailed balance, at least not with respect to the correct physical entropy production.  One can sometimes introduce effective parameters, like an effective temperature to maintain an intuition but that seems more aligned with a conservative approach as alluded at in the introduction of the present notes; see e.g. \cite{2nd,stef,tim}.\\

The idea in general is the following.  The probe perturbs the stationary condition of the medium by moving in it.  The perturbation is felt as a change in the interaction potential $U(q,x) \rightarrow U(q,x) + (q_s-q)\,V(x)$ with potential $V(q,x) = \partial_qU(q,x)$.  As a consequence, the medium responds following formul{\ae} like \eqref{6n}.  For example, suppose first that the medium is in fact in equilibrium.  Taking the amplitude $h_s=q_s-q_t$ we get a response on the expected force $\langle V(q_t,x_t) \rangle$ as
\[
{\cal K}= \int_0^t \id s \, (q_s-q_t) \frac{\id}{\id s}\langle V(q_t,x_s) V(q_t,x_t)\rangle_\text{eq} 
\]
by following formula \eqref{kuboa}. Partial integration in time yields the usual friction term as in a generalized Langevin equation, with friction kernel related to the noise, having a covariance in terms of force-force correlations.  That would give the usual Einstein relation. But for a nonequilibrium medium that responds to the probe motion we must replace \eqref{kuboa} with \eqref{6n} and an additional term. the frenetic contribution indeed, will need to be added to $\cal K$, while the noise formula remains essentially unchanged.  We refer to \cite{2nd,stef, tim} for more details.

\section{Frenetic bounds to dissipation rates}

Thermalization or relaxation to equilibrium refers to the property of reduced systems or variables of showing convergence in time to an equilibrium condition or value as set by the constraints or conserved quantities in the larger system.  The phrasing ``reduced system'' either refers to a collection of macroscopic variables or empirical averages, e.g., the spatial profile of some mass or energy density, or to some set of local observables e.g. belonging to a subvolume.  The idea is that many other degrees of freedom are left in contact with the reduced system, are integrated out, and constitute a dynamical heat bath for the degrees of freedom of the reduced system.  That is also why the derivation of Brownian motion or of more general stochastic evolutions in an appropriate coupling- and scaling-limit is an essential ingredient in studies of thermalization.  The literature on the subject is vast and has been in the forefront of statistical physics up from the time of Boltzmann till today where the relaxation of quantum systems remains a hot topic.  A recurrent question there has been how to unify the idea of a unitary or Hamiltonian dynamics with a possibly dissipative dynamics for the reduced system.  Often ergodic properties of the dynamics have been called upon to make time-averages to coincide with some ensemble averages etc.  We will not discuss that issue here.  As a more original contribution we emphasize here that the structure of gradient flow as shown in macroscopic evolution equations is in fact pointing to two separate aspects of relaxation, one which is entropic and dominates close-to-equilibrium, and one which is frenetic and is crucial also far-from-equilibrium.  We use here the word ``frenetic'' to have a complement to ``entropic'' that emphasizes the component of dynamical activity in relaxation processes.  Secondly, we give a framework, mainly dynamical fluctuation theory, in which both relaxation and stationarity can be discussed and we indicate how to obtain from there frenetic bounds to the dissipation rate. \\

Gradient flow gives a differential-geometric characterization of certain dissipative relaxational processes for macroscopic systems.  The heuristics is as follows.  Suppose the reduced system finds itself in some condition $X$, cf. the introduction \ref{int}.  The latter can refer to a specific spatial profile of some density, in which case we really have a function $\rho(r), r\in V$ over some spatial domain $V$, or to the value of some macroscopic variable like the averaged magnetization.  The condition $X$ need not be the one of equilibrium compatible with the present constraints, and we ask in what direction the system will evolve from $M$.  That displacement is given by the change in $X$, or e.g. for a density profile $\rho$ in terms of a current $j^*(\rho)$ which is one of many possible currents compatible with $\rho$ and with any further constraints on the system.  The question is what determines that $j^*$ from $\rho$.  The answer has two parts, one thermodynamic and the other kinetic.  The thermodynamic part refers to the maximizing of the total entropy or the minimizing of some free energy.  Gradient flow makes the corresponding thermodynamic potential a Lyapunov function, monotone in time; that is part of the H-theorem in the case of the Boltzmann equation for the thermalization of dilute gases.  So gradient flow moves in the direction of lower free energy or larger entropy for the total universe,  which is the usual Boltzmann picture of selecting that trajectory which moves over macroscopic conditions that show ever larger phase space volumes to end up finally in the equilibrium condition where the phase space volume is huge compared to the initial ones. But that is not enough; moving out and in macroscopic conditions requires certain amounts of accessibility and escape-ability -- after all, there is a good reason that relaxation times might be large.  The point is that gradient flow is also following {\it steepest descent} of the free energy.  That steepest descent requires a distance or metric in the space of macroscopic conditions; we must know the height lines of the free energy and where the gradient is largest from the given $M$.  That distance is not provided by the thermodynamics but by the kinetics of the process.  What enters for standard dissipative evolutions of a density field $\rho$ is the Onsager matrix of mobility or the diffusion constant, as function of the condition $\rho$.  The positivity of that Onsager matrix provides the local measure of distance (the metric). All that is rather well-studied and known but less so in the nonlinear regime where the condition is still far-from-equilibrium, and where the linearity between the thermodynamic force (gradient of the potential) and the current does not need to hold.

\begin{figure}[h]
\centering
\includegraphics[width=7 cm]{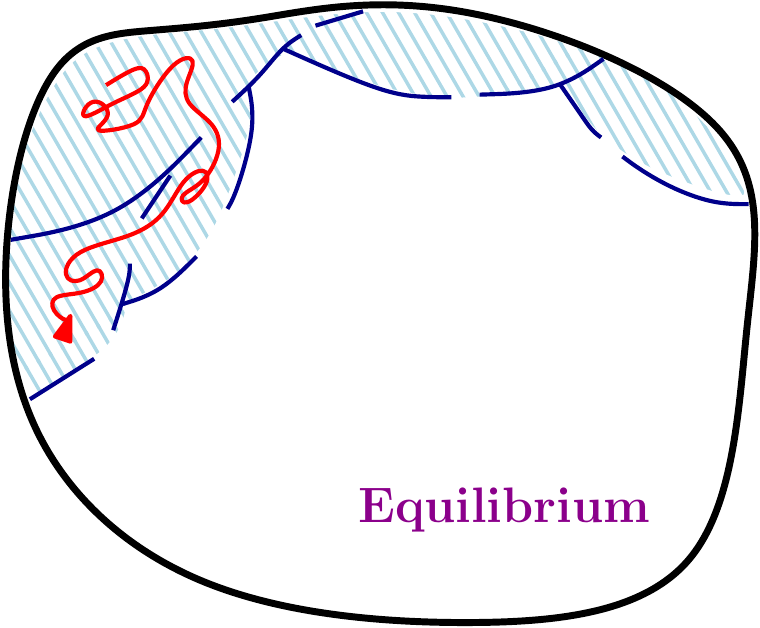}
\caption{The Boltzmann picture of relaxation to equilibrium, amended by the influence of ``surface effects'' of the phase space region.  Higher dynamical activity or escape rates on the reduced or coarsed grained level of description decreases relaxation times.}\label{phsp}
\end{figure}

Concerning the main point of these lectures indeed, it should be clear that non-dissipative aspects should play a role also.  Especially when starting well away from equilibrium, visiting phase space regions with smaller volume, it appears not unreasonable to imagine that also ``surface'' effects and not only volume effects will contribute to determine the pathways to equilibrium.  The exit and entrance rates from a macroscopic condition are not determined by its volume (entropy) solely.  That is the first heuristics behind suspecting that time-symmetric quantities, like time-symmetric currents, will also appear essentially outside equilibrium.\\
In fact, on the same level of discussion we can easily imagine that the interaction between various components lead to effective kinetic constraints such that the relaxation becomes very difficult.  That ``breaking of ergodicity'' is studied in various models and sometimes is referred to as glassy behavior or as many body localization \cite{gar}.  The kinetic constraints are basically involving non-dissipative aspects, having to do with trapping and caging effects in phase space, which is just an abstraction of what happens in Fig.~\ref{traj} (replacing there the external field by thermodynamic forces and replacing the obstacles by either disorder in the interaction or by specific many body effects).\\

Let us finally come also here to a more mathematical analysis.  As often it pays to consider the problem on space-time and to go on the level of fluctuation theory.  That is the whole idea of dynamical statistical ensembles, as we saw at work e.g. in Section \ref{reps} on response.  Let us however not start this time from more microscopic models but imagine how the macroscopic equation for the relaxation to equilibrium, or more generally the displacements or currents would be derived as function of the given macroscopic condition.  We take therefore the view that we have a family of {\it possible} trajectories $((\rho_s,j_s), s\in [0,t])$ over a time-interval $[0,t]$ starting at time zero from the condition $\rho_0$.  The latter has also a certain probability which is the subject of static fluctuations we touched in Sections \ref{onst} and \ref{vari}.  Indeed the $\rho_s$ are the possible density profiles on a certain level of description, so we would have e.g. $\rho_s(x)$ for the probability or density of particles at or in state $x$.  The $j_s$ are the corresponding currents or displacements, and they are often constrained in the sense that only those trajectories have non-zero probability for which 
\begin{equation}\label{dyfl}
\dot{\rho_s} + Dj_s=0
\end{equation}
where the operator $D$ could  be a (discrete or continuum) divergence in the case of a conserved quantity, or also just unity for a reactive dynamics.  Apart from such constraints as in \eqref{dyfl} all is possible; dynamical fluctuation theory is supposed to give us information about the probabilities.  These probabilities satisfy a law of large numbers in the limit where a scaling factor $N$ tends to infinity;  $N$ can indicate the size of the system or the number of particles etc.; the precise origin and context for \eqref{dyfl} can vary wildly.  Fluctuations are corrections to the law of large numbers and we assume therefore the asymptotic in $N$ form,
\begin{equation}\label{lo}
\text{Prob}[(\rho_s,j_s), s\in [0,t]] = e^{N{\cal S}[\rho_0]}\,\,\exp[-N\int_0^t L(\rho_s,j_s)\, \id s]
\end{equation}
in terms of the static fluctuation rate ${\cal S}$ (called, nonequilibrium statistical entropy) and the Lagrangian $L\geq 0$.  Via that scale $N$  we can assume that there is a unique most typical trajectory starting from $\rho_0$.  In other words we can find for every density (condition) $\rho$ what comes next by finding $j^* = j^*(\rho)$ as solution of the zero-cost flow:
\begin{equation}\label{zcf}
L(\rho,j^*(\rho) =0
\end{equation}
All other possibilities are exponentially small in $N\uparrow \infty$.  If we are dealing with relaxation processes, the solution $j^*(\rho)$ will determine the macroscopic evolution equation by plugging it in for the current in \eqref{dyfl}. One example where $j^*(\rho)$ turns out to be linear in $\rho$ is the Master equation for Markov processes. In case we are dealing with a stationary nonequilibrium process where the density is $\rho_s=\rho^*$ constant in time, we obtain through $j^*(\rho^*)$ information about the currents maintained in the system.  Whatever the case  we may in particular be interested in the expected entropy production rate
\begin{equation}\label{sr}
\sigma(\rho) = \sigma(\rho,j^*(\rho))
\end{equation}
corresponding to $\rho$.  It is the dissipation rate, product of forces and fluxes, expressed as function of the present system condition $\rho$.  In the case of thermalization (relaxation to equilibrium for a system open to energy exchanges at fixed environment temperature) \eqref{sr} is minus the rate of free energy decrease when in $\rho$. Lower bounds on $\sigma(\rho)$ would give information about relaxation times. In stationary nonequilibrium (density $\rho^*$) then $\sigma^* = \sigma(\rho^*,j^*(\rho^*))$ is the mean (stationary) entropy production rate.  Then, lower bounds on $\sigma^*$ would give refinements on the second law like statement that $\sigma^*\geq 0$. To have a closer look we can see to relate that entropy production rate to the dynamical fluctuations in \eqref{lo}.\\
The way \eqref{sr} is related to the Lagrangian (and the dynamical fluctuations) is, as always, via the condition of local detailed balance \eqref{ldb}.  More precisely, the entropy production rate corresponding to a pair $(\rho,j)$ of density and current, is the bilinear form that will be obtained from the part in the Lagrangian that is antisymmetric under time-reversal:
\[
\sigma(\rho,j) = L(\rho,-j) - L(\rho,j)
\]
(We ignore here the presence of currents for odd degrees of motion, like momentum currents.)
The non-dissipative aspect of the Lagrangian (and dynamical fluctuations) resides in the 
time-symmetric part, which we split up further in two parts:
\[
L(\rho,j) + L(\rho,-j) = 2\psi(\rho,j) + 2L(\rho,0)
\]
For the interpretation we can look at specific models, but here is already some structure.  Suppose we have a density $\rho$ and no current whatsoever, $j_s=0$ for all times.  Then, it must be that $\rho_s=\rho_0$ did not change, and the probability of such a trajectory 
is according to \eqref{lo} proportional with
\[
\exp -N t L(\rho_0,0)
\]
In other words, $L(\rho,0)$ is the escape rate from $\rho$. Its inverse gives us the information of the mean time the system resides in $\rho$, indeed invariant under time-reversal, but of course depending on $\rho$ and on all possible forces.  Secondly, the $\psi(\rho,j)$ is symmetric in $j$ and non-zero when $j\neq 0$.  It is therefore the analogue of the activity parameters that we first introduced under \eqref{ps}.  It depends of course also on $\rho$ and it gives the unoriented traffic in $\rho$ associated to $j$.\\

We are ready for the final line in the reasoning.  If we look back at the zero-cost flow \eqref{zcf}] with solution $j^*(\rho)$, we obviously have that
\begin{equation}\label{frene}
\frac 1{2}\,\sigma(\rho,j^*(\rho)) = \psi(\rho,j^*(\rho)) + L(\rho,0)
\end{equation}
which says that the expected current $j^*(\rho)$ given $\rho$ is such that the dissipation rate (left-hand side) equals the expected dynamical activity (right-hand side). We have here the analogue of \eqref{9n} but for macroscopic densities. The identity \eqref{frene} has consequences in terms of frenetic bounds for the expected entropy production rate, cf. \cite{frenb}.  Any lower bound on the right-hand side of \eqref{frene} provides a lower bound on the expected dissipation. Thus we will get lower bounds on the mean entropy production rate which are symmetric in the currents and non-dissipative, which leads to refinements of the second law version which merely says that  $\sigma(\rho,j^*(\rho)) \geq 0$. It is important here to note that $\psi(\rho,j)$ in convex in $j$ whenever $L(\rho,j)$ is convex in $j$; that is because the time-antisymmetric part $\sigma(\rho,j)$ is linear in $j$ to be compatible with local detailed balance. Because by construction $\psi(\rho,0) =0$ and $\psi(\rho,j)=\psi(\rho,-j)$ we then have that $\psi(\rho,j) > 0$ from the moment $j\neq 0$.  So from the moment there is a non-zero entropy production, we will have already the non-trivial bound $\sigma^* \geq \psi(\rho^*,j^*(\rho^*))>0$.  Finer frenetic bounds are discussed in \cite{frenb}.

\section{Symmetry breaking}
We finally come to a major influence of non-dissipative aspects on nonequilibrium structures.  To put it in contradictory terms and in line with the recent \cite{jack}: purely dissipative trajectory ensembles are wrong because they do not allow dissipation.
By a  purely dissipative trajectory ensemble we mean a probability distribution on trajectories, as in Section \ref{reps}, where however the action with respect to equilibrium only contains the entropy flux.  In other words, looking back at \eqref{sta} and \eqref{deco}, we would have an action ${\cal A} = \lambda \,S$ proportional to the entropy flux only, thus forgetting about the frenetic contribution $D$. Such ensembles (which, disrepectfully, we will call {\it mutilated}) are often considered for modeling nonequilibrium in the spirit of maximum entropy or maximum caliber assumptions\footnote{More positively, such maximum entropy principles can teach us what observables or quantities are missing in the variational formulation.  Here for example we will learn that there is more than the values of currents or of entropy flux alone that characterize the nonequilibrium ensemble.}. Yet, the point is that these mutilated ensembles
\begin{equation}\label{mut}
\text{Prob}[\omega] \propto \text{Prob}_\text{eq}[\omega]\,e^{\lambda S(\omega)}
\end{equation}
often have additional symmetries which are unwanted. The main mathematical consequence of \eqref{mut} is that under time-reversal $\theta$ (defined in Section \ref{path}), $S\theta=-S$ and hence,
\[
\text{Prob}[\theta\omega] \propto \text{Prob}_\text{eq}[\omega]\,e^{-\lambda S(\omega)}
\]
by the time-reversal symmetry of the equilibrium ensemble.
There may however still be other involutions $\Gamma$ leaving invariant the equilibrium ensemble for which $S(\Gamma\omega) = -S(\omega)$ and hence, for the mutilated ensemble \eqref{mut},
\[
\text{Prob}[\Gamma\theta\omega] = \text{Prob}[\omega] 
\]
(see equation (12) in \cite{jack} or equations (6)--(9) in \cite{mar}).
Indeed, it is easy to imagine for example how spatial reflection of $\omega$ inverts $S(\omega)$ in exactly the same way as does time-reversal (and leaving the equilibrium reference invariant). It implies that we could exchange time-reversal with spatial reflection, or more generally that we could induce time-reversal by reversing the thermodynamic forces.  Then, the dynamical ensemble of  a diffusion driven by some non-conservative field $\cal E$ would have a time-reversal which is obtained by simply flipping $\cal E \rightarrow -\cal E$. Now, since the stationary (single time) distribution of the process is identical to the stationary distribution of the time-reversed process, we would have that the stationary distribution is left invariant by flipping the field $\cal E \rightarrow -\cal E$, which of course is often very wrong.  But even forgetting about that last point\footnote{After all, the idea of stationary distribution has lost most of its meaning now since the trajectories distributed by \eqref{mut} are not associated to a Markovian time-evolution.}, all expectations of time-symmetric observables would be invariant under $\cal E \rightarrow -\cal E$, $\langle \Gamma O\rangle = \langle O\rangle$ when $\theta O = O$, and all expectations of time-antisymmetric observables would be antisymmetric  under $\cal E \rightarrow -\cal E$, $\langle \Gamma O\rangle = - \langle O\rangle$ when $\theta O = -O$.  For example, from the last equality, for an energy current $J$ we would have $\langle \Gamma J\rangle = - \langle J\rangle$ but often $\Gamma J = J$ so that we would need to conclude that $\langle J\rangle =0$, no dissipation!\\

 In other words, such nonequilibrium ensemble \eqref{mut}, lacking non-dissipative terms in the action, would not be able to break certain symmetries.  That point was first stressed in the theory of \cite{mar}.  Also an example was given there of a viscous fluid under influence of a pressure difference in some tube.   In the laminar flow regime (left part of Fig.~\ref{summ}) we indeed cannot distinguish between reversal
of time and reversal of pressure difference.  Yet, at higher pressure differences (right part of Fig.~\ref{summ}), when things get really nonequilibrium, the laminar pattern breaks up, the flow becomes turbulent and time-reversal definitely differs from field-reversal.  Another and older example where such symmetry breaking is essential is found in \cite{kroy}.  There a turbulent boundary layer flow is considered over a smooth symmetric hump; see Fig.~\ref{summ}.
\begin{figure}[thb]
  \centering
  \includegraphics[width=14cm]{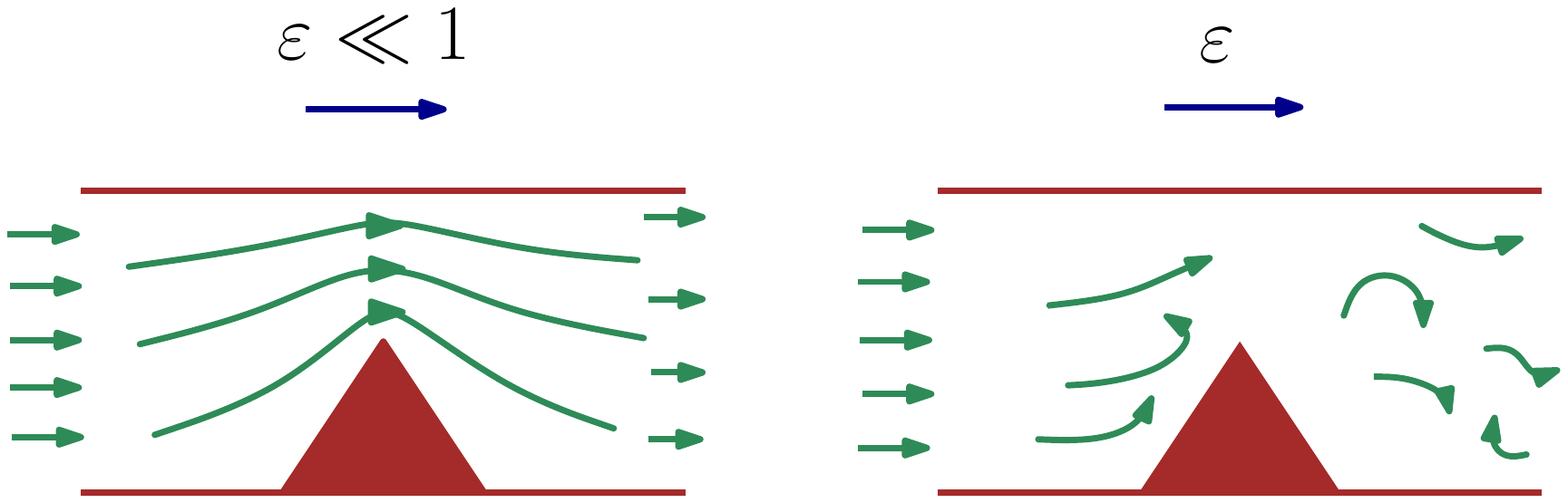}
  \caption{ Flow of particles driven by a field ${\cal E}$. Left:  for small driving we get a laminar flow without symmetry breaking.  Time-reversal equals field reversal.  Right: For appreciable field strength turbulent flow develops and the spatial symmetry gets broken essentially by the non-dissipative  frenetic contribution in the action, which is absent in the ensemble \eqref{mut}.}
  \label{summ}
\end{figure}
  For the ensemble \eqref{mut} there would never be a possibility to break that symmetry.  Yet, true nonequilibrium does create a weak symmetry breaking, indeed visible for aeolian sand transport and dune formation.

Note still that the ensemble \eqref{mut} is compatible with the Kubo formula and linear response around equilibrium, exactly for the (same) reasons 
that lead to \eqref{2n4} and \eqref{gk}.  Moreover, so called fluctuation symmetries \cite{poincare} are trivially true.  Yet, they lack the non-dissipative elements that are essential for creating interesting nonequilibria.  To maintain the paradoxical phrasing, interesting dissipative structures crucially depend on non-dissipative aspects.

\section{Frenometry}
The question arises how non-dissipative aspects can be isolated and/or measured.  Is there something like measuring excesses in dynamical activity, as we encounter for example in the frenetic contribution to response functions? Obviously all effects discussed in these notes can be inverted: whenever we understand how activity parameters have an influence, then we can also start from that theory to learn about these activity parameters from the observed facts. Let us first see what that means for thermodynamic potentials.\\
Entropy is of course not directly measured --- it is obtained via measurements of heat capacity or from heat measurements more generally.  Similarly, free energies are measured from isothermal work, as we remember from the following type of calculation.  Suppose
 that the states $x$ of a system are coupled to some real variable $q$, say the position of a probe, via the interaction energies $E(x,q)$.
The mechanical force on $q$ is $-\nabla_q E(x,q)$.  For a quasi-static probe we can take the statistical average over the equilibrium distribution,
\[
f_\text{eq}(q) = \sum_x \frac 1{Z_q}e^{-\beta E(x,q)}\left(-\nabla_q E(x,q)\right) = \frac 1{\beta}\nabla_q\log Z_q = -\nabla_q {\cal F}(q)
\]
In other words, the statistical force is gradient with potential given by the free energy (now depending on $q$).  It is then clear how reversible isothermal transformations will have changes in free energy given by the work done on the probe.\\
The idea for measuring non-dissipative aspects can proceed similarly.\\

In general we have that statistical forces out-of-equilibrium are given by
\[
f(q) = \sum_x \rho_q(x)\,\left(-\nabla_q E(x,q)\right)
\]
where $\rho_q(x)$ is the stationary distribution of a nonequilibrium medium with degrees of freedom $x$, which is coupled via interaction energy $E(x,q)$ to a quasi-static probe with position $q$.  Clearly, as we have seen throughout these notes, the distribution $\rho_q(x)$ will contain non-dissipative aspects. 
 Therefore, measuring the work done by the statistical force $f(q)$ on a probe will reveal aspects of dynamical activity, changes in escape rates or in time-symmetric parts of reactivities.  That is the general idea, but of course we need to implement it in specific situations.  Paradoxically, as the system dissipates more, more non-dissipative features become visible.

\subsection{Reactivities, escape rates}
Look again at the situation depicted in Fig.~\ref{traj}.\\
Suppose you would like to measure the dependence of the escape rate $\psi(W,x)$ on the external driving field.  What we will do is to couple the walker to a probe with position $q$ for example as a load connected to the walker via a spring,
\[
E(x,q) =\frac{\lambda}{2} (x-q)^2
\]
The stationary distribution of the driven particles for high field $W$ can be approximated by dividing an elementary interval length $L$, say between two major obstacles, in three different states, one for the walker being directly behind the obstacle, another for being in the middle of the interval and yet a third state for being in front of an obstacle; see Fig.~\ref{traj}.  We then write $\rho_q(x) \simeq z/\psi(W,x)$, independent of probe position $q$ for very small $\lambda$.  That is similar to what we saw already in Fig.~\ref{3st}, how the reactivities determine the stationary distribution.
The statistical force on the probe (load) is then approximated by
\[
f(q;W,\lambda) = z\,\lambda \int_0^L (q-x)\,\frac 1{\psi(W,x)}\,\id x
\]
telling us about the escape rate $\psi(W,x)$. When we find the position of the probe where the statistical force is zero, $f(q^*;W) =0$, we will get information about the escape rate profile. For example, if we take $\psi(W,x) = e^{-aWx}, x\in [0,L]$, then 
the stationary position of the probe will be at $q^* = L - 1/(aW) + \ldots$ for large $W$, from which we would find the slope coefficient $a$.  The probe will find itself in a harmonic well for effective potential $\lambda(q-L + 1/(aW))^2/2$.

\subsection{Non-gradient aspects are non-dissipative}

Another tool to derive information about non-dissipative aspects of a nonequilibrium medium is to look at the rotational part in the statistical force induced on a probe.  In fact, the very presence of a rotational component to the statistical force is produced by the simultaneous appearance of excess entropy flux and excess dynamical activity when the medium relaxes to a new stationary condition after displacing the probe.\\

The simplest example of how the statistical force can pick up the dependence of reactivities on the driving goes as follows; see also \cite{prl}.\\
We follow a probe on the unit circle, $q\in S^1$, in contact with a thermal bath at inverse temperature $\beta=1$, modeled via the overdamped Langevin dynamics
\[
\gamma\dot{q} = -\frac{\partial}{\partial q} E(x,q) + \sqrt{2\gamma}~ \xi_t
\]
for $\xi_t$ standard white noise, $\gamma$ is the damping coefficient and  there is an interaction with an `internal' degree of freedom $x=-1,0,1$ with potential
\[
E(x,q) = x \sin q + 2 x^2 \cos q \label{eq:Ux}
\]
The $x$ is the fast degree of freedom, driven with transition rates
\[ 
k^q(x,x')= e^{-\frac{\beta}2 [E(x',q)-E(x,q)]} \,\phi(x,x')\, e^{\frac{1}{2} s(x,x')}
\] 
The drive is uniform, with $s(-1,1)= s(1,0)= s(0,-1) = \varepsilon$ but we assume that also the symmetric activity parameters $\phi(x,x') = \phi(x',x)$  are affected via
$\phi(-1,1) = \phi(1,-1)= \phi_0(1 + a |\varepsilon|)$ for some $a\geq 0$, while
$\phi(0,\pm 1)=\phi(\pm 1,0)=1$.  Detailed balance is achieved at $\ve=0$, where $\phi_0$ picks up the relative importance of the dynamical activity over the transitions $1\leftrightarrow -1$.

The statistical force on the probe is
\begin{equation}\label{fq}
f(q) = -\langle x \rangle^q \cos q + 2 \langle x^2 \rangle^q \sin q
\end{equation}
and can be calculated exactly. For driving $\varepsilon \neq 0$ there are no symmetries forbidding the probe to have net motion (Curie's principle) and hence to show a systematic current round the circle.  The rotational part of the force
$f_\text{rot} = \oint f(q)\,\id q$ is  plotted  versus $\varepsilon$ in Fig.~\ref{rotf}(a).  Observe that the rotational force depends on the coefficient $a$, and hence picks up information about that non-dissipative part $\phi(x,x')$ in the reaction rates.  In fact, we see clearly, take $a=1/2$, that the rotational force is maximal for an intermediate $\varepsilon \simeq 4$.  For larger $a\geq 0$ we get less rotational force on the probe, because the medium gets more jammed, in fact similarly again to what happens in Fig.~\ref{3st}(b) for large $b$.\\  

\begin{figure}[thb]
  \centering
  \includegraphics[width=14cm]{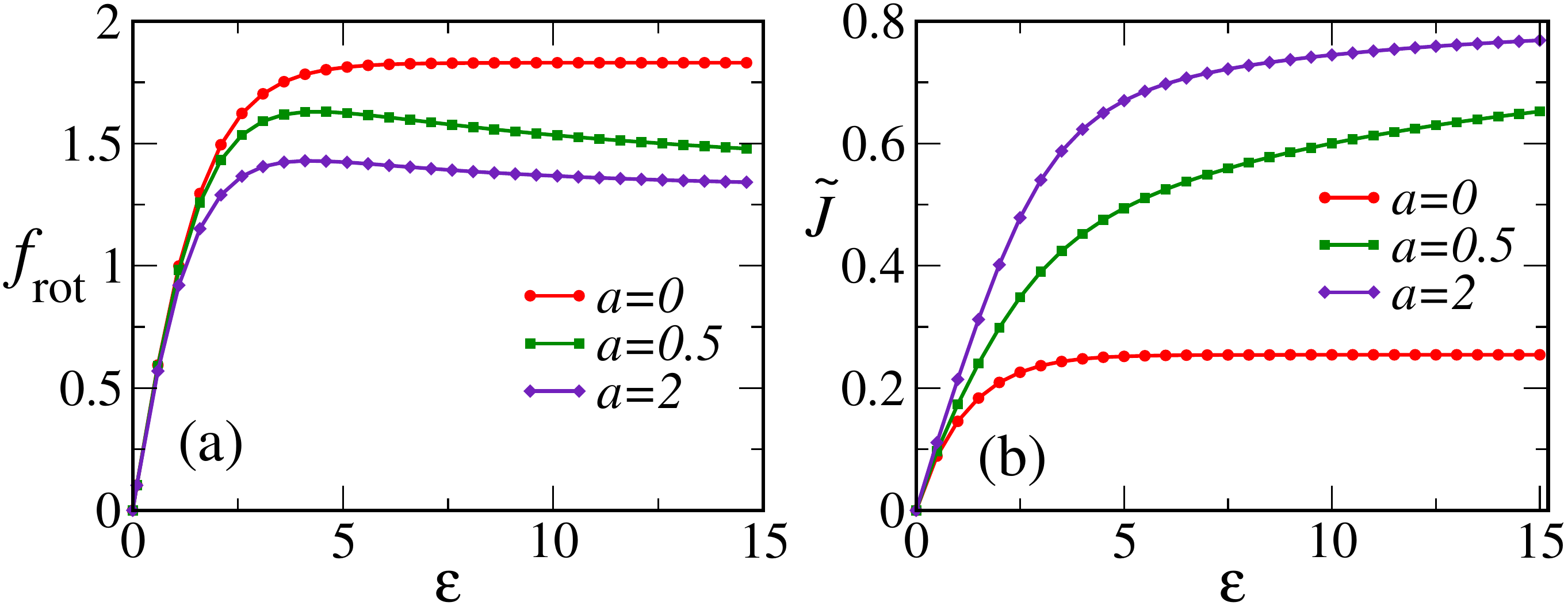}
  \caption{(a) The rotational part $f_\text{rot} = \oint f(q)\,\id q$ of the statistical force \eqref{fq} on the probe and how it depends on the excess in time-symmetric reactivities for intermediate values of $\varepsilon$ only. (b) The current in the driven three-state process rescaled by $\exp\varepsilon/2$.  In all cases the dependence on activity parameter $a$ is a second-order effect in $\varepsilon$.}
  \label{rotf}
\end{figure}

Fig.~\ref{rotf}(b), where we see the current in the medium process, must be compared with Fig.~\ref{rotf}(a):  there is no simple relation between the rotational power in the probe and the entropy production in the medium.  Non-dissipative players decide when well away from equilibrium ($\varepsilon=0$).  We see that in the dependence on the parameter $a$ for the change in reactivity with $\varepsilon$ but the time-symmetric part in the action is of course also changing with $\varepsilon$.

\section{Conclusions}
Non-dissipative aspects of nonequilibirum systems are crucial for a number of phenomena ranging from the dependence of the stationary distribution on activity parameters to the role of time-symmetric path-variables in response and fluctuation theory.  Escape rates, reactivities and undirected traffic may all be modified by the driving and that excess plays a role in response and relaxation theory.  What is called dynamical activity or the frenetic contribution complements entropic/dissipative arguments for the physical understanding of the nature of the stationary distribution, for determining the direction of currents, for nonlinear response behavior and for the emergence of so called dissipative structures. Surely there is much more to life processes than dissipation and irreversible or stochastic thermodynamics will fail for essential characteristics of nonequilibria when not naming the role of dynamical time-symmetric activity \cite{life}. It remains to be seen how phase transitions and indeed pattern formation gets influenced by non-dissipative effects under nonequilibrium conditions.  That brings us to further challenges in the construction of nonequilibrium statistical mechanics.  One can see that nonequilibrium is so diverse because of the non-thermodynamic, kinetic non-dissipative aspects that shape it.  A useful construction will therefore be able to identify a minimal set of variables measuring dynamical activity.\\

\noindent {\bf Acknowledgment}:  These notes originate in lectures given at the Winter school on Complexity at the TU Eindhoven in December 2015. Special thanks to Mark Peletier for the encouragement to start writing them down.  I am also very indebted to my collaborators of which the joint papers appear below in the references.



\begin{thebibliography}{10}
\bibitem{Groot} S. De Groot, and P. O. Mazur, {\it Non-Equilibrium Thermodynamics}, Dover Pub. (1962).

\bibitem{mmcs}
C.~Maes, What decides the direction of a current?  Mathematics and Mechanics of Complex Systems {\bf 3}, 275--295 (2016).


\bibitem{kls}
S.~Katz, J.L.~Lebowitz, and H.~Spohn, Stationary nonequilibrium states for stochastic lattice gas models of
ionic superconductors, 
J. Stat. Phys. {\bf 34}, 497 (1984).

\bibitem{time} C.~Maes and K.~Neto\v{c}n\'y, 
Time-reversal and Entropy. 
J. Stat. Phys. {\bf 110}, 269 (2003).



\bibitem{har} 
T.~Harada and S.-I.~Sasa,  Equality connecting energy dissipation with violation of fluctuation-response relation.
Phys. Rev. Lett. {\bf 95}, 130602 (2005).


\bibitem{der} B.~Derrida, Non-equilibrium steady states: fluctuations and large deviations of the density and of the current.
J. Stat. Mech. P07023 (2007).


\bibitem{hal} H.~Tasaki, Two theorems that relate discrete stochastic processes to microscopic mechanics.
{\tt arXiv:0706.1032v1}. 

\bibitem{cor}
{\tt http://www.nil.wustl.edu/labs/corbetta/research\_spont\_activity.html}. Last visited 1 June 2017.


\bibitem{luit}
F.~Michler, H.~van Hees, D.D.~Dietrich, S.~Leupold and  C.~Greiner, Non-equilibrium photon production arising from the chiral mass shift.
{\tt 	arXiv:1304.4093 [nucl-th]}.


\bibitem{schEP}
C.~Maes and K.~Neto\v{c}n\'{y}, Minimum entropy production principle. Scholarpedia {\bf 8}(7), 9664 (2013).

\bibitem{physA}
C.~Maes and W.~O'Kelly de Galway, On the kinetics that moves Myosin V. Physica A: Statistical Mechanics and its Applications {\bf 436}, 678--685 (2015).

\bibitem{woj}
W.~De Roeck and C. Maes, Symmetries of the ratchet current. Phys. Rev. E {\bf 76}, 051117 (2007).

\bibitem{pieter}
P.~Baerts, U.~Basu, C.~Maes and S.~Safaverdi, The frenetic origin of negative differential response, Phys. Rev. E {\bf 88}, 052109 (2013).

\bibitem{zia}
R.K.P.~Zia, E. L.~Pr{\ae}stgaard, and O.G.~Mouritsen, Getting more from pushing less: Negative specific heat and conductivity in nonequilibrium steady states. Am. J. Phys. {\bf 70}, 384 (2002).

\bibitem{urna}
U.~Basu and C.~Maes, Mobility transition in a dynamic environment. J. Phys. A: Math. Theor. {\bf 47}, 255003 (2014).

\bibitem{sarrac}
O.~B\'enichou, P. Illien, G. Oshanin, A. Sarracino, R. Voituriez, Microscopic theory for negative differential mobility in crowded environments.
Phys. Rev. Lett. {\bf 113}, 268002 (2014).

\bibitem{bsv}
M.~Baiesi, A.~Stella, and C.~Vanderzande, Role of trapping and crowding as sources of negative differential mobility. Phys. Rev. E{\bf 92}, 042121 (2015).


\bibitem{db} 
M.~Barma and D.~Dhar, Directed diffusion in a percolation network. J. Phys. C {\bf 16}, 1451 (1983).

\bibitem{sol}
F.~Solomon, Random walks in a random environment. Ann. Prob. {\bf 3}, 1--31 (1975).


\bibitem{thi}
T.~Demaerel and C.~Maes, Activity induced first order transition for the current in a disordered medium.
{\tt  	arXiv:1705.02820 [cond-mat.stat-mech]}.  To appear in Condensed Matter Physics.

\bibitem{scholar}
C.~Rouvas-Nicolis and G.~Nicolis, Stochastic resonance. Scholarpedia, 2(11):1474 (2007).


\bibitem{chan}
R.~Jack, J.P.~Garrahan, and D.~Chandler,
Space-time thermodynamics and subsystem observables in kinetically constrained models of glassy materials.
J. Chem. Phys. {\bf 125}, 184509 (2006).
 
 \bibitem{gar}
J.~M.~Hickey, S.~Genway, J.P.~Garrahan, Signatures of many-body localisation in a system without disorder and the relation to a glass transition. {\tt arXiv:1405.5780v1}. 

\bibitem{Japan} R. Kubo, K. Matsuo, K. Kitahara, Fluctuation and relaxation of macrovariables. J. Stat. Phys. {\bf 9}, 5--95 (1973).

\bibitem{macro} L.~Bertini, A.~De Sole, D.~Gabrielli, G.~Jona-Lasinio, and C.~Landim, Macroscopic fluctuation theory. Rev. Mod. Phys. {\bf 87}, 593 (2015).

\bibitem{derr} B.~Derrida, J.L. Lebowitz, E.R. Speer, Large deviation of the density drofile in the steady state of the symmetric simple exclusion process. J. Stat. Phys. {\bf 107}, 599--634 (2002).

\bibitem{einstein} A.~Einstein, Theorie der Opaleszens von homogenen Fl\"ussigkeiten und Fl\"ussigkeitsgemischen in der N\"ahe des kritischen Zustandes. Annalen der Physik {\bf 33}, 1275 (1910).

\bibitem{lanford} 
O.~E.~Lanford, Entropy and equilibrium states in classical statistical mechanics. {\it Statistical Mechanics and Mathematical Problems}, Springer Lecture Notes {\bf 20}, 1-113, 1973.

\bibitem{martinlof}
A. Martin L\"of, {\it Statistical Mechanics and the Foundations of Thermodynamics},
Volume 101 Lecture Notes in Physics Series, Springer-Verlag, 1979.

\bibitem{ellis}
R. Ellis, {\it Entropy, Large Deviations, and Statistical Mechanics}, Springer, 2006.

\bibitem{matrixprod}
B.~Derrida, M.R. Evans, V. Hakim, V. Pasquier, Exact solution of a 1d asymmetric exclusion model using a matrix formulation. J.Phys. A{\bf 26}, 1493--1518 (1993).

\bibitem{lowT}
C.~Maes, K.~Neto\v{c}n\'{y} and W.~O'Kelly de Galway, Low temperature behavior of nonequilibrium multilevel systems. J. Phys. A: Math. Theor. {\bf 47}, 035002 (2014).


\bibitem{win}
C.~Maes and W.~O'Kelly de Galway, A low temperature analysis of the boundary driven Kawasaki Process. J. Stat. Phys. {\bf 53}, 991--1007 (2013).

\bibitem{lincu}
S.~Bruers, C.~Maes and K.~Neto\v{c}n\'{y}, On the validity of entropy production principles for linear electrical circuits. J. Stat. Phys. {\bf 129}, 725--740 (2007).


\bibitem{kolk}
U.~Basu and C.~Maes, Nonequilibrium Response and Frenesy. J. Phys.: Conf. Ser. {\bf 638}, 012001 (2015).

\bibitem{land}
R.~Landauer, Inadequacy of entropy and entropy derivatives in characterizing
the steady state. Phys. Rev. A. {\bf 12}, 636--638 (1975).

\bibitem{heatbounds}
C.~Maes and K.~Neto\v{c}n\'{y}, Heat bounds and the blowtorch theorem. Ann. H. Poincar\'e {\bf 14}, 1193--1202 (2013).

\bibitem{hopf}
J.J.~Hopfield, Kinetic Proofreading: A New Mechanism for Reducing Errors in Biosynthetic Processes Requiring High Specificity. Proc. Nat. Acad. Sci. {\bf 71}, 4135--4139 (1974).

\bibitem{frey}
S.~N.~Weber, C.~A.~Weber and E.~Frey, Binary Mixtures of Particles with Different Diffusivities Demix. Phys. Rev. Let. {\bf 116}, 058301 (2016).

\bibitem{ioanny}
A.~Y.~Grosberg, J.-F.~Joanny, Nonequilibrium statistical mechanics of mixtures of particles in contact with different thermostats.
Phys. Rev E{\bf 92}, 032118 (2015). 

\bibitem{solon}
A.~P.~Solon, Y.~Fily, A.~Baskaran, M. E.~Cates, Y. Kafri, M.~Kardar, and J.~Tailleur,
Pressure is not a state function for generic active fluids.
Nature Physics {\bf 11}, 673--678 (2015).

\bibitem{poincare} C. Maes, On the origin and the use of fluctuation relations for the entropy. In: J. Dalibard, B. Duplantier, V. Rivasseau (Eds.), S\'eminaire
Poincar\'e, vol. 2, Birkhäuser, Basel, 2003, pp. 29–62.


\bibitem{roy1}
C.~Maes, S.~Safaverdi, P.~Visco and F.~van Wijland, Fluctuation-response relations for nonequilibrium diffusions with memory. Physical Review E {\bf 87}, 022125 (2013).

\bibitem{roy2}
M.~Baiesi, C.~Maes and B.~Wynants, The modified Sutherland-Einstein relation for diffusive non-equilibria. Proc. Royal Soc. A 467, 2792--2809 (2011).


\bibitem{fdr}
M.~Baiesi, C.~Maes and B.~Wynants, Fluctuations and response of nonequilibrium states. Phys. Rev. Lett. {\bf 103}, 010602 (2009).

\bibitem{pccp} U.~Basu, M.~Kr\"uger, A.~Lazarescu, and C.~Maes, Frenetic aspects of second order response.  Phys. Chem. Chem. Phys. {\bf 17}, 6653 (2015).

\bibitem{beh}
L.~Helden, U.~Basu, M.~Kr\"uger and  C.~Bechinger, Measurement of second-order response without perturbation.  EPL {\bf 16}, 60003 (2016).


\bibitem{harrisevans}
E.~Levine, D.~Mukamel and G.M.~Sch\"utz, Zero-Range Process with Open Boundaries. J. Stat. Phys. {\bf 120}, 759--778 (2005).


\bibitem{2nd}
C.~Maes, On the Second Fluctuation-Dissipation Theorem for Nonequilibrium Baths. J. Stat. Phys. {\bf 154}, 705--722 (2014)

\bibitem{stef}
C.~Maes and S.~Steffenoni, Friction and noise for a probe in a nonequilibrium fluid.  Phys. Rev. E {\bf 91}, 022128 (2015).

\bibitem{tim}
C.~Maes and T.~Thiery, The induced motion of a probe coupled to a bath with random resettings. {\tt arXiv:1705.07670 [cond-mat.stat-mech]}.

\bibitem{prl}
U. Basu, C. Maes, K. Neto\v{c}n\'{y}, How Statistical Forces Depend on the Thermodynamics and Kinetics of Driven Media.  Phys. Rev. Lett. {\bf 114}, 250601 (2015).

\bibitem{frenb}
C.~Maes, Frenetic bounds on the entropy production. {\tt arXiv:1705.07412 [cond-mat.stat-mech]}.

\bibitem{jack}
R.L.~Jack and R.M.L.~Evans, Absence of dissipation in trajectory ensembles biased by currents. {\tt arXiv:1602.03815v1 [cond-mat.stat-mech]}.

\bibitem{mar}
C.~Maes and M.H.~van Wieren, Time-symmetric fluctuations in nonequilibrium systems. Phys. Rev. Lett. {\bf 96}, 240601 (2006).

\bibitem{kroy} 
G.~Sauermann, K.~Kroy and H.J.~Herrmann, Continuum saltation model for sand dunes.  Phys. Rev. E{\bf 64}, 031305 (2001).

\bibitem{life}
M.~Baiesi and C.~Maes, Life efficiency does not always increase with the dissipation rate. In preparation.


\end{thebibliography}
\end{document}